# Investigating the relationship between biodiversity and ecosystem multifunctionality: Challenges and solutions

Jarrett E. K. Byrnes[1*], Lars Gamfeldt[2], Forest Isbell[3], Jonathan S. Lefcheck[4], John N. Griffin[5], Andrew Hector[6], Bradley J. Cardinale[7], David U. Hooper[8], Laura E. Dee[9], J. Emmett Duffy[4]

1 – National Center for Ecological Analysis and Synthesis, 735 State Street, Santa Barbara, CA 93101, USA
2 – Department of Biological and Environmental Sciences, University of Gothenburg, Box 461, SE-40530, Gothenburg, Sweden
3 – Department of Ecology, Evolution, and Behavior, University of Minnesota, St Paul, Minnesota 55108, USA
4 – Virginia Institute of Marine Science, College of William & Mary, Gloucester Point, VA 23062, USA
5 – Department of Biosciences, Swansea University, Singleton Park, Swansea, SA2 8PP, UK
6 – Department of Plant Sciences, University of Oxford, South Parks Road, Oxford, OX1 3RB, UK
7 – School of Natural Resources & Environment. University of Michigan. Ann Arbor, MI 48109
8 – Department of Biology, Western Washington University, Bellingham, WA 98225-9160, USA
9 – Bren School of Environmental Science and Management, University of California, Santa Barbara, CA 93106, USA
* - Current Address: Department of Biology, University of Massachusetts Boston, 100 Morissey Blvd., Boston, MA 02125, USA

## Abstract

Extensive research shows that more species-rich assemblages are generally more productive and efficient in resource use than comparable assemblages with fewer species. But the question of how diversity simultaneously affects the wide variety of ecological functions that ecosystems perform remains relatively understudied, and it presents several analytical and empirical challenges that remain unresolved. In particular, researchers have developed several disparate metrics to quantify multifunctionality, each characterizing different aspects of the concept, and each with pros and cons. We compare four approaches to characterizing multifunctionality and its dependence on biodiversity, quantifying 1) magnitudes of multiple individual functions separately, 2) the extent to which different species promote different functions, 3) the average level of a suite of functions, and 4) the number of functions that simultaneously exceed a critical threshold. We illustrate each approach using data from the pan-European BIODEPTH experiment and the R *multifunc* package developed for this purpose, evaluate the strengths and weaknesses of each approach, and implement several methodological improvements. We conclude that a extension of the fourth approach that systematically explores all possible threshold values provides the most comprehensive description of multifunctionality to date. We outline this method and recommend its use in future research.

*"You keep using that word. I do not think it means what you think it means." – I. Montoya, 1987*

## INTRODUCTION

Nearly 20 years of empirical work has clearly shown that losing species can impact a wide variety of ecosystem processes such as primary production and nutrient cycling (Balvanera *et al.* 2006; Cardinale *et al.* 2006; 2011), and that these impacts may equal or exceed those of many other human drivers of environmental change (Hooper et al. 2012; Tilman et al. 2012). These experiments primarily focus on biodiversity's effect on single functions. However, accumulated evidence suggests that the impact of diversity is different, and potentially stronger, when multiple functions are considered together (Hector & Bagchi 2007; Gamfeldt et al. 2008). Here we consider the growth and development of research on biodiversity and multiple ecosystem function, and how we can best evaluate how diversity simultaneously can affect ecosystem 'multifunctionality'.

Most experiments to date have measured the impacts of diversity loss on one or a few functions considered in isolation (see summaries by Hooper *et al.* 2005; Stachowicz *et al.* 2007; Cardinale *et al.* 2011). For such individual ecosystem processes, effects of diversity generally saturate at relatively low levels of species richness (see data summaries by Cardinale *et al.* 2006; 2011; but see Reich *et al.* 2012). In practice, human society values a suite of ecosystem properties, each of which has potential to respond to diversity loss (e.g., MEA 2005). It would clearly be valuable to quantify how ongoing diversity loss simultaneously influences the suite of functions or services that ecosystems provide, and whether the effect of diversity on multiple functions is different from the effect on individual functions. Our current understanding of how diversity affects ecosystem functioning may be limited or even biased by the current single-function approach if trade-offs or synergies among processes are ignored.

A few empirical studies suggest that diversity may increase the provision of several ecosystem processes simultaneously—the so-called 'multifunctionality' of ecosystems—and that effects of diversity on multifunctionality may not saturate at the low levels typical of single functions (e.g., Duffy *et al.* 2003; Hector & Bagchi 2007). Thus the magnitude of diversity's impact may be stronger when multifunctionality is considered. Alternatively, trade-offs among different functions could render diverse systems less capable of providing multiple functions compared to monocultures of particular species (Zavaleta et al. 2010, Gamfeldt et al. 2013). The effect of diversity on multifunctionality thus could be smaller than the effect on any single function. However, we cannot assess the strength of diversity's effect on multifunctionality from extant work because the few experiments that considered how diversity affects multiple functions simultaneously have used multiple analytical frameworks to measure multifunctionality.

While we can define multifunctionality as the simultaneous performance of multiple functions, how this definition is operationalized makes a critical difference to the conclusions drawn from an experiment. Researchers have used four basic approaches to explore the relationship between biodiversity and multifunctionality (Table 1). We briefly present and then discuss them in more detail below. The simplest is *the single* *functions approach*, which considers a collection of functions and asks qualitatively whether more functions achieve higher values in the diverse mixture than at lower levels of species richness (Duffy et al. 2003). Analysis of these univariate responses provides information about the diversity-multifunctionality relationship but does not provide any quantitative measure of multifunctionality. A second, related method (Hector and Bagchi 2007, Isbel et al. 2011), the *turnover approach*, tests whether different sets of species promote different functions and has potential to quantify the fraction of species that contribute to one or more functions. Third, the *averaging approach* (Hooper and Vitousek 1998) aims to collapse multifunctionality into a single metric that estimates the average value of multiple functions achieved in a given assemblage or plot. Fourth, the *threshold approach* (Gamfeldt et al. 2008, Zavaleta et al. 2010) tallies the number of functions that quantitatively exceed some pre-defined threshold of "functionality" in a given assemblage or plot.

These four approaches provide very different means of evaluating the relationship between diversity and multiple ecosystem functions and they require different assumptions and interpretations. Each has pros and cons (Table 1). None provides a single omnibus metric of multifunctionality as currently implemented. Moreover, all approaches share issues that require consideration in estimating multifunctionality. For example, an inherent challenge in estimating multifunctionality is deciding whether a negative or positive value of a function is considered 'desirable'. This decision is necessary to create a single number as an index of multifunctionality. It is also inherently subjective and requires an explicit explanation of the rationale.

In this paper, we provide a critical analysis of the four existing approaches for measuring multifunctionality. We demonstrate the insights provided by modified versions of each and we compare their strengths and weaknesses. We illustrate each technique using the R package *multifunc* (http://github.com/jebyrnes/multifunc; installation instructions code for this paper are in Supplementary Information 1) applied to data from the European BIODEPTH experiment (Spehn et al. 2005), a series of simultaneous experiments that manipulated diversity of grassland plants at eight locations across Europe. Ultimately, we conclude that a modified version of the threshold approach provides the most comprehensive and informative approach and recommend its use for future research. Our hope is that this analysis will pave the way for more rigorous and consistent analyses of the influence of biodiversity (or other factors) on ecosystem multifunctionality.

## MEASURING MULTIFUNCTIONALITY

### First steps: the single functions approach

A simple first pass at examining whether species diversity influences multifunctionality is to qualitatively compare how diversity affects each of a group of functions individually. Does a diverse mixture increase the levels of multiple functions beyond what the average or even the best performing single species does? Duffy et al. (2003) first took this approach that amounts to a qualitative precursor of the threshold approach we describe later. That study examined how functions in seagrass mesocosms responded to manipulation of grazer species richness. Most functions reached their highest levels in single-species treatments, consistent with a sampling effect. However, different grazers maximized (achieved the highest measured values of) different functions, such that only mixtures of grazer species achieved similarly high values of several functions simultaneously. Thus, Duffy et al. concluded that multiple species were necessary to support multiple functions simultaneously at high levels (what they called the “multivariate dominance effect”). In such cases, where data for all monocultures are available, a simple examination of single functions can provide clues as to how diversity influences multifunctionality.

*An example of the single functions approach: the German BIODEPTH site*

To demonstrate the application of the single function approach and provide an introduction to the dataset used throughout this paper, let us consider the pooled data from both blocks of year three at the German site of the pan-European BIODEPTH experiment (see Spehn et al. 2005 for details). Briefly, BIODEPTH was a suite of parallel experiments in which eight grasslands spanning the continent of Europe were seeded with different numbers of plant species drawn from local species pools. After three years, seven functions were measured: aboveground biomass, belowground biomass, cotton decomposition, wood decomposition, light penetration, soil nitrogen, and plant nitrogen, although not all functions were measured at all sites. In Germany, a subset of five ecosystem functions were measured: aboveground biomass production, belowground biomass, cotton decomposition, soil nitrogen, and plant nitrogen. Here we follow the original authors and consider greater decomposition rates and lower soil nitrogen as desirable, the latter because it indicates greater total resource use.

To examine the results for single functions, we fit separate linear models to estimate how each of the five functions changes with species richness, rather than comparing levels of functions in diverse communities to monoculture values (Figure 1; original analyses were on a log scale, but the qualitative results do not differ). We found clear effects of species richness on two functions (aboveground biomass $F_{1,58}$=35.91, p<0.0001; and total plant nitrogen (g N/m$^2$) $F_{1,58}$=15.25, p<0.001), some support for richness effects on two others (cotton decomposition $F_{1,58}$=2.91, p=0.09; and soil inorganic nitrogen pools sizes [nitrate + ammonium] $F_{1,58}$=3.15, p=0.08), and no effect of richness on belowground biomass ($F_{1,58}$=0.03, p=0.86). In all cases but root biomass, the trend was for species richness to increase function. In total, it appears that diversity enhances multifunctionality in Germany by increasing the levels of performance of more than one function (aboveground biomass and soil nitrogen). But given the weak relationships between diversity and two of the functions, this strength of this conclusion is ambiguous. Looking at those functions for which we have data, a single species, the legume *Trifolium pratense*, had the highest observed values for four of the five functions, further calling this conclusion into question (Spehn et al. 2005).

*Strengths and weaknesses of the single functions approach*

Should researchers perform additional analyses of relationships between diversity and multifunctionality, the single functions approach helps illuminate which individual processes drive trends in multifunctionality. In the cases where there is a strong, consistent positive or negative relationship between diversity and all measured functions, a researcher may not need to perform additional analyses to argue that biodiversity affects multifunctionality.

The single functions approach, however, cannot tell a researcher quantitatively how diversity affects multifunctionality.  It only provides information for a qualitative statement.  Furthermore, if some results are negative or nonlinear, even a qualitative statement about multifunctionality may not be straightforward.  Thus, the single functions approach cannot be used as a standalone assessment of multifunctionality.

**The Turnover Approach: Do different sets of species affect different functions?**

A positive link between species richness and multifunctionality depends on either variation among species in their contributions to different functions and/or interactions among species that enhance multiple functions. In cases where the single functions approach suggests evidence for multifunctionality, each function still may be driven by one or a few species (the selection effect *sensu* Aarssen 1997; Loreau & Hector 2001). The question remains whether there is "turnover with respect to multifunctionality", i.e., whether different species are responsible for different functions.

Estimating turnover in functional contributions requires determining which species contribute to each function and then assessing the redundancy of species contributions to each function (Hector & Bagchi 2007). This involves two steps, each of which requires careful consideration.

First, we must quantify the effect of each species on each ecosystem function by modeling the level of a given function in each plot as a function of the presence/absence of each species. The data requirements for such models can be high: generally they require good replication of each individual species in both monoculture and multiple different species mixtures across a diversity gradient. Previous studies have used linear models without interactions (Hector & Bagchi 2007; He et al. 2009; Isbell et al. 2011), though allowing for interactions may more accurately capture species' contributions to functions (Kirwan et al. 2009). We discuss when this is appropriate, and the additional issues raised by models with interaction effects below. Another open question is how to implement appropriate model selection techniques to identify species that contribute significantly to a function. While this question is particularly relevant to the overlap approach, it is beyond the scope of our discussion here.

Second, we obtain the relationship between the number of functions and the cumulative number of species influencing at least one function by examining the overlap (i.e., as quantified by a similarity index) of the contributing species pools for each function or combination of functions from fit models. The estimate of species turnover among different functions is the inverse of this overlap.

Turnover can be evaluated quantitatively by examining how the cumulative number of species influencing ecosystem functioning changes with the number of functions, via simulation of all possible combinations of functions. At one extreme, if each species uniquely influences one and only one function, then the number of species promoting ecosystem function would scale positively and linearly with the number of functions with a slope of 1 (Hector and Bagchi 2007). At the other extreme, if all ecosystem processes were influenced by the same set of species, the slope would be zero. Previously observed relationships lie between these two extremes, indicating that there is some turnover in species between functions (Hector and Bagchi 2007; Isbell et al. 2011).

Critically, the relationship between species richness and multifunctionality depends on the relative proportion of positive and negative effects of species and the sizes of those effects. To date the turnover approach has only been applied to species with a positive effect on an ecosystem function (Hector & Bagchi 2007; He et al. 2009; Isbell et al. 2011). However, some species could also have negative effects on some functions; including such effects can have large implications for how diversity influences ecosystem multifunctionality.

To integrate the overlapping effects of both positive and negative contributions, we propose two additional analyses. First, we suggest a test for correlation between the number of significantly positive and significantly negative effects of a species on different ecosystem functions. This estimates the potential for trade-offs among functions and whether species have predominantly positive or negative effects on ecosystem functions. Second, investigators can estimate the impact of trade-offs more quantitatively by examining the relationship between the average standardized positive effect size of a species on all functions it affects and the average standardized negative effect size of a species on all functions it affects, drawn from the statistical fits developed in the first step, above. By standardized, we refer to standardized regression coefficients where $r_{xy} = b * sd_x/sd_y$. This indicates whether species have quantitatively similar positive and negative effects on those functions they impact. By combining these two relationships, we can infer whether positive and negative effects cancel each other out, or whether one type is dominant. For example, if species have predominantly positive effects on different functions, and the strengths of positive effects are greater on average than those of negative effects, then turnover of species likely contributes to a positive relationship between biodiversity and ecosystem multifunctionality.

*An example of the turnover approach*

Returning to the BIODEPTH example from Germany, we applied a stepwise AIC model selection approach to fit linear additive models to each function to obtain the minimally adequate set of species affecting each function, as in Hector and Bagchi (2007). We then examined the relationship between the number of functions and the cumulative fraction of the species pool that had a positive or negative effect on those functions (Fig. 2). As the number of functions considered increases, a larger cumulative fraction of species had a positive effect on at least one of the functions in the set. The relationship appears to approach saturation as the number of functions increases (i.e., at 5 functions). For the average single function, roughly 19.4% of the planted species pool (note, other analyses have used observed species pool) had a significant positive effect and 14.8% had a negative effect. When all five functions were considered, roughly 54% of species contributed positively and 48% contributed negatively to the set of functions. These two results suggest that there is substantial functional uniqueness among the species that affect the functions in the BIODEPTH Germany site – both positively and negatively.

The net effect of diversity on multifunctionality depends on the balance of positive and negative effects by species. We found that as the number of positive effects that a species had on ecosystem functions increased, so did the number of negative effects (Fig. 3a, Spearman rank correlation = 0.47, $t_{1,29}$=1.93 p=0.006). Similarly, as positive effect sizes increased, so did negative effect sizes, suggesting tradeoffs of species effects on different ecosystem functions (correlation between the average positive and negative effect sizes of species: Spearman rank correlation = -0.75, $t_{1,29}$=-4.14, p<0.001). However, effect sizes of positive effects tended to be larger than effect sizes of negative effects (Fig. 3b, deviation from a 1:-1 line, t=2.96, df=30, p=0.0059), suggesting that as more species are added, there will be a net gain in multifunctionality. Note, however, that the deviation here is driven largely by one species, *Crepis biennis*. With that species removed, the deviation of positive and negative effect sizes from the 1:-1 line becomes nonsignificant (p=0.691), indicating no net effect of adding more species to the community. Thus, it is difficult to say from the overlap analysis alone whether diversity is strongly linked to actual levels of multifunctionality in the German BIODEPTH site.

*Strengths and weaknesses of the turnover approach*

The turnover approach provides a way to evaluate whether different sets of species drive different functions. The biggest strength of this approach is that it identifies which species have positive, negative, or neutral effects for each function and tests whether these sets differ among functions. By generalizing this approach to examine the balance of positive and negative effects, we can more accurately determine how different functions depend on different species, and whether such differences can explain observed effects of diversity on multifunctionality.

The turnover approach has several shortcomings, however. It provides no quantitative estimate of the extent to which changes in species richness influence multifunctionality. If this approach unambiguously demonstrates greater positive than negative species' effect sizes across species and functions, then increasing diversity will increase multifunctionality in the system. But the turnover approach does not measure multifunctionality *per se*.

Furthermore, the turnover approach has two stringent data and analytic requirements necessary for interpretation. First, the relationships between diversity and individual functions should have similar sign and form; if they differ strongly, results from the turnover approach can be difficult to interpret. Second, estimating individual species effects requires designs that include each species in a variety of compositional treatments so that variable selection techniques can determine which species are important. Since many current biodiversity-ecosystem function experiments consist solely of monocultures and a single mixture treatment, they may not be amenable to analysis using the turnover technique. Even when data are available to estimate effect sizes of individual species, if species interact (e.g., the effect of species A changes in the presence of species B, as through competition or facilitation), the modeled effect sizes could be inaccurate. Models that include such interactions may better estimate species' effect sizes, but also require even more data (Kirwan et al. 2009). For these reasons, we recommend that researchers exercise caution when and where they utilize the turnover approach.

**The averaging approach: What is the average effect of changing diversity on multiple ecosystem functions?**

A simple technique for summarizing ecosystem multifunctionality involves averaging standardized values of multiple functions into a single index. This averaging approach was first suggested as an index of 'relative resource use' (Hooper & Vitousek 1998) to summarize the depletion of multiple types of nutrients by a plant assemblage for comparison across plots with different richness of plant functional groups. This averaging approach represents apparently the first attempt at a measurement of multifunctionality in the diversity-function literature. Its simplicity has led others to use it as well (Mouillot *et al.* 2011; Maestre *et al.* 2012a; b).

The general application of the technique is straightforward. For each function measured, standardize the values to have the same scale. For functions where negative values indicate higher levels of function (e.g., low soil nitrate equates to high resource use by plants), values should be 'reflected' to appear on the same scale (see below) before standardizing. An index of average function is then created by taking the mean value across all functions in a plot. This averaged multifunctionality index ($MF_a$) for a plot can be expressed as

$$MF_a = \frac{1}{F}\sum_{i=1}^{F} g\left(r_i\left(f_i\right)\right) \quad (1)$$

where F is the number of functions being measured, $f_i$ are the measures of function i, $r_i$ is a mathematical function that reflects $f_i$ to be positive, if deemed necessary (see discussion in the introduction), and g is a transformation to standardize all measures of function to the same scale. For functions that need to be reflected, $r_i$ can take the form of $-f_i+\max(f_i)$ or just $-f_i$, depending on considerations based on standardization functions discussed below. More complex types of averaged indices (e.g., taking into account variances, geometric means, etc.) are of course also possible.

Once any necessary reflections have been done so that all functions are in the desired direction, the values for each function must be standardized before averaging to remove the effects of differences in measurement scale between functions. There are a wide variety of standardization methods, all of which yield similar results (Maestre *et al.* 2012b). The two most common are the z-transformation (Mouillot *et al.* 2011; Maestre *et al.* 2012b) and a standardization by a maximum observed value (Hooper & Vitousek 1998; Maestre *et al.* 2012a). While z-transforming functions may improve their properties for analysis using traditional linear statistics (Maestre *et al.* 2012b), we recommend standardizing by a chosen maximum value as we have done here. We make this recommendation both because 1) researchers may want to move beyond linear models and 2) standardizing the scale of a multifunctional index by a maximum value creates a metric that is intuitively interpretable: the proportion of maximum multifunction achieved by a single plot. We recognize that using a maximum value for standardization has some problems (e.g., sensitivity to outliers). See the thresholding approach below for several alternative ways of determining a maximum.

After standardizing the measured values of each function within an individual experimental plot, the standardized values are averaged to obtain a mean value of functioning for that plot. Note that equation 1 assumes that all functions are weighted equally in calculating the averaged multifunction index. Alternate weightings are also possible, which may be desirable in applied management. For example, if biomass production is deemed twice as important as decomposition for land management, a weighted averaged index may be preferable. In either case, the index can be used to assess how the average level of multiple ecosystem functions changes with diversity.

*An example of the averaging approach*

First, we identify those functions from the German BIODEPTH data for which lower values are considered positive contributions to ecosystem functioning, and reflect them by multiplying by -1 (equation 1). We reflected values of soil nitrogen, as low values indicate greater resource capture or greater mobilization of N from the soils despite loss to leaching (Scherer-Lorenzen et al. 2003). We also added the unreflected maximum to this function so that the lowest level of transformed function is 0 as discussed above. We then divided each function by its maximum value to scale all functions between 0 and 1. Finally, we created the multifunctionality index by taking the unweighted average of all five functions.

A linear model shows that plant richness positively affects average function, $MF_a$ at the German BIODEPTH site (Fig. 4, Diversity $F_{1,22}$=22.554, p<0.001, slope estimate = 0.0133 ± 0.0028 SE). This is interpreted to mean that for every species added to the system, the average value across all functions increases by roughly 1% of its maximum value. This averaging approach suggests that species richness increases multifunctionality in the German site. However, without combining this approach with the single function analysis, we cannot say whether this result is driven by diversity affecting one, two, or all of the observed functions. Thus there is still substantial ambiguity as to whether this is a representative measure of multifunctionality on its own.

*Strengths and weaknesses of the averaging approach*

The averaging approach provides a seemingly intuitive way to assess changes in several ecosystem functions simultaneously. Its interpretation—change in the average level of a suite of ecosystem functions—is clear. Very high levels of the $MF_a$ index (e.g., near 1) unambiguously mean that many functions are achieving high levels of performance. But the averaging approach only provides clearly interpretable results at high values of $MF_a$. In such cases, most functions must be performing at high levels. At intermediate values of $MF_a$, however, it is not possible to distinguish between multiple functions performing at intermediate values from some performing at high values while others perform at low values. Although the averaged index values are equal in the two cases, we would interpret them in very different ways with respect to multifunctionality. Second, at very low values of $MF_a$, we cannot distinguish the case where a treatment has no effect on either of two functions from the case where diversity has a positive effect on one function and a negative effect on the other. Thus it is necessary to look at both the single function and $MF_a$ curves separately to discern the underlying relationship between biodiversity and multifunctionality. Finally, in practical terms, we may often not view different functions as substitutes, meaning that a decrease in one function cannot be compensated by an increase in another (Gamfeldt et al. 2008).

**The Threshold approach: Are multiple species needed to maximize multiple functions?**

To remedy the weaknesses of the averaging approach, we need to evaluate whether multiple functions are simultaneously performing at high levels. This is accomplished by the threshold approach. In the biodiversity-multifunctionality literature, previous efforts to accomplish this have tallied the number of functions that simultaneously surpass some threshold (Gamfeldt & Hillebrand 2008; Zavaleta *et al.* 2010; Peter *et al.* 2011), generally by creating an index of the number of functions surpassing the threshold in each experimental plot or unit (Zavaleta et al. 2010).

To calculate the threshold-based index of multifunctionality, $MF_t$, one first needs to define a threshold. This threshold is normally some percentage of the maximum observed value of each function. Other biologically or management related reference values can also be used. Different thresholds can even be used for different functions. The maximum approach raises two questions. First, what value should be used as the 'maximum' for an experiment? Second, what is the appropriate percentage of that maximum? For the first question, the highest observed value for a function could be taken as an estimate of the highest attainable value, but because it is necessarily a single observation, it could also be an outlier due to observation error, process noise, or other factors. The chance of using a spuriously high maximum value can be reduced by averaging multiple values to estimate the maximum. Here, we use the mean of the n+1 highest measurements of a function across all richness levels in an experiment as our maximum, where n is the smallest sample size of a single richness treatment level. An alternative approach is to define the maximum by some arbitrary subset, say the top 5%, of all plots (Zavaleta et al. 2010) or some relevant management target.

Once we have settled on a maximum value for each function, we next must decide on the appropriate proportion of that maximum value for each function to serve as our threshold value, $t_i$, to create a threshold index for a given plot:

$$MF_t = \sum_{i=1}^{F} \left( r_i(f)_i > t_i \right) \quad (2)$$

where F is the total number of functions and $f_i$ is the value for function i in a given plot, which may be reflected as discussed in previous sections (using the mathematical function $r_i$). Note that, as with the averaging approach, functions may need to be reflected to appropriately capture the desired direction of effects. From this equation we can see that the choice of threshold will influence the value of the resulting index, and we anticipate that diversity will have a stronger association with multifunctionality at some choices of threshold than at others. We return to this issue in the section on the multiple threshold approach below.

*An example of the threshold approach*

For the five functions measured at the German BIODEPTH site, we assessed multifunctionality at an arbitrarily chosen threshold of 80% of each function, using as the maximum value the mean of the seven highest observations for each function (n=6 for the 16 species polyculture). We first calculated the number of functions performing at or above 80% of this maximum in each plot. We then fit a generalized linear model with a quasipoisson error to estimate a linear relationship predicting the number of functions performing at or above their threshold as a function of planted species richness. We selected this model after considering a number of issues regarding using count data and the model's functional form, both of which can be influenced by the goals of the analysis and the experimental design (Supplementary Material 2) and can differ between experiments.

We evaluated the fit of our model regressing multifunctionality on species richness in two ways. First, we performed a likelihood ratio test to show that the inclusion of species richness provided a better fit than a model with only an intercept ($\chi^2$=15.91, DF=1, p<0.001). Second, we estimated the coefficient describing the relationship between species richness and multifunctionality (the number of functions reaching at least 80% of maximum) as 0.113 ± 0.033 SE, which is strongly supported as being different from zero (t=3.42, p=0.001). This coefficient estimate means, roughly, that ten additional species are needed to bring one more function above our chosen threshold at the Germany site.

Does the strength of the diversity effect on multifunctionality change with the choice of threshold? One might expect that diversity has a stronger effect as higher thresholds are imposed, i.e., that more species are needed to maintain a suite of functions at high thresholds. To evaluate this question we first calculated $MF_t$ (equation 2) using 20%, 40%, 60%, and 80% as our threshold values. Diversity positively affected the number of functions exceeding threshold at values of 40 and 60%. In contrast, the relationship between richness and $MF_t$ became flatter at higher threshold values and the intercept was lower (Fig. 5), where few functions exceeded the threshold at any level of diversity. The relationship was also relatively flat at low threshold values but the intercept was higher, where nearly all functions achieved the threshold. The threshold approach shows that diversity influences multifunctionality at the German BIODEPTH site, but that the strength of this relationship is sensitive to the choice of threshold value.

*Strengths and weaknesses of the threshold approach*

Assessing multifunctionality at the plot level with a threshold-based approach provides a powerful, flexible method. It captures the number of functions performing well even in the presence of tradeoffs and correlations among functions. For example, if one function is always maximized when another is minimized, then this tradeoff will be clear because the number of functions greater than a threshold will never equal the total number of functions measured. The threshold approach can also be used whether the relationship between diversity and individual functions is linear or nonlinear.

The coefficient describing the relationship between diversity and number of functions reaching a threshold has a clear interpretation. If a linear function is fitted, the slope represents the change in number of functions meeting threshold per species added or subtracted. If, on the other hand, the relationship is exponential (i.e., $\log(MF_t)$ is a linear function of richness), then the slope can be easily transformed: $e^{richness}-1$ is approximately the proportional change in function per change in number of species.

Last, in the unusual event that a researcher can measure every unique function of interest in a system (for example, in a specific management application), logistic regression can be used to estimate a coefficient that is interpreted as the log odds ratios of including a new function per species added or subtracted. If only some are measured, the relationship may not be asymptotic within the bounds of the number of functions measured.

Despite these advantages, the single threshold approach is not a perfect measure of multifunctionality for at least three reasons. First, the choice of threshold is often arbitrary. Second, the magnitude of each function is captured imperfectly and only indirectly as the threshold changes. For example, even if multiple functions pass a threshold value, $MF_t$ does not reveal whether they pass by a small or large margin. Finally, examining only a single threshold value may miss some critical value at which diversity has its strongest impact. Conversely, choosing a threshold based on diversity's strongest effect involves circular reasoning, and should be avoided. We lose information with this approach either way.

## The Multiple Threshold Approach to Evaluating Diversity Effects on Multifunctionality

If the choice of an arbitrary threshold can obscure diversity's role in influencing multifunctionality, what is the solution? The changing slope of $MF_t$ on species richness at different thresholds (Fig. 6) suggests a solution to the problem of arbitrary thresholds and a way of more fully examining the fingerprint of diversity on multifunctionality. This involves plotting the effect of diversity on multifunctionality ($MF_t$) across the full range of thresholds between 0 and 100%.

A systematic examination of this distribution provides an information-rich picture of how diversity influences multifunctionality. Plotting threshold choice (x) against slope of $MF_t$ on richness (y) reveals multiple pieces of information about how diversity influences multifunctionality, not limited to just the maximum effect of diversity (slope) and the threshold at which this effect is achieved (Fig. 7). The curves in Figures 6 and 7 provide a profile of the effect of diversity on multifunctionality. These curves provide several key metrics that can help us understand the relationship between diversity and multifunctionality via examining multiple threshold choices. Four metrics in particular give us key information about the how diversity can influence multifunctionality:

- **Minimum Threshold ($T_{min}$):** The lowest threshold where diversity begins to have an effect (i.e., a slope that is significantly greater than or less than 0). This indicates the proportion of maximal functioning at which multifunctionality becomes influenced by changes in species richness.

- **Maximum Threshold ($T_{max}$):** The value of the threshold beyond which the slope first declines to be not significantly different from zero. This measures the upper threshold beyond which diversity has no effect on multifunctionality.

- **Threshold of Maximum Diversity Effect ($T_{mde}$):** The value of the threshold where diversity has its strongest positive or negative effect (i.e., most extreme slope values on the y axis), Fig. 8d-f.

- **Realized Maximum Effect of Diversity ($R_{mde}$):** The strength of the relationship (i.e., slope) where diversity has its strongest positive and/or negative effects indicates the maximum observed effect size of species richness on number of functions surpassing the threshold.

We can conclude that diversity is a strong driver of multifunctionality if $T_{min}$ is low, $T_{max}$ is high, and both Tmde and $R_{mde}$ are likewise high. These metrics do not tell the complete story, of course. Additional metrics can provide more nuance to our interpretation of how biodiversity influences multifunctionality. This nuance may be ideal for examining how biodiversity influences multifunctionality in a single system, but less informative when comparing across systems and experiments. These metrics are:

- **Minimum Diversity Independent Multifunction ($M_{min}$):** The number of functions achieving the threshold at $T_{min}$. In combination with $T_{min}$ this indicates whether, independent of diversity, the system has low or high baseline multifunctionality, and thus how much influence diversity can have relative to a baseline.

- **Maximum Diversity Independent Multifunctionality ($M_{max}$):** The number of functions achieving the threshold at $T_{max}$. This measures the number of functions that are able to achieve high levels of performance in a system simultaneously.

- **Diversity Maximized Multifunctionality ($M_{mde}$):** The number of functions achieving $T_{mde}$ at the highest level of diversity.

- **Proportion of maximum possible diversity effect ($P_{mde}$):** The slope of $R_{mde}$ can be compared with the maximum possible slope of the relationship for the design of the experiment, which is simply the # of functions divided by highest number of species used in the experiment. $P_{mde}$ gives the proportion of the maximum possible relative importance of diversity for multifunctionality realized in this experimental system.

*An example of the multiple threshold approach*

Examining the relationship between threshold choice and slope of the diversity- $MF_t$ relationship shows that in Germany, diversity had a moderate impact on multifunctionality (Fig. 6, Table 2). The 95% confidence intervals around the estimated slopes reveal whether the estimates overlap 0, giving a test of the threshold values at which diversity has no effect on multifunctionality. For Germany, the relationship peaks at a threshold of 50% ($T_{mde}$) with a slope of roughly 0.17 functions added per species. However, species richness is positively related to multifunctionality ($MF_t$) at thresholds between roughly 15% ($T_{min}$) and 98% ($T_{max}$). For a sense of scale of the strongest effect of diversity, we note that for the German

BIODEPTH experiment's range of diversity levels (1-16 species) and number of functions measured (5), the maximum possible slope of the relationship between species richness and the number of functions greater than a threshold is ~ 0.312 (i.e. 5/16). At its strongest, diversity thus has 54% ($P_{mde}$) of the maximum possible effect on multifunctionality within the design of this experiment. Diversity could not simultaneously drive all functions to their maxima; that is, while diversity did increase multifunctionality in this system above values seen in monocultures, and had significant positive effects even at high threshold values (90-95%), the decreasing slope at higher thresholds indicates that high species richness did not guarantee that all functions performed at their highest levels. Indeed, from Fig. 6, $M_{max}$ ~ 1 species. Thus, diversity has a moderate effect on multifunctionality based on our criteria above.

We can further illustrate the value of the multiple threshold approach by comparing metrics of diversity effects on multifunctionality across sites in a comparative context. Here, we compare three other BIODEPTH sites: Portugal, Sweden, and Sheffield (Fig. 8, Table 2) as they show three contrasting patterns. In Portugal, diversity had a small positive effect on multifunctionality ($MF_t$) at low threshold values but none at moderate to high thresholds. This means that increasing diversity could promote multifunctionality only if low thresholds of functioning were sufficient. The effects of diversity on multifunctionality in Portugal are weak. In Sweden, by contrast, diversity drove multiple functions to moderate levels at low to medium thresholds, but had no effect at high thresholds ($T_{max}$ = 73%). $P_{mde}$ was roughly equivalent to Germany (53%) as was $M_{max}$ (~2.2 species) although at a lower threshold. Thus, Sweden has a moderate effect of diversity on multifunctionality, similar to Germany. Finally, at Sheffield, diversity had a clear positive effect on multifunctionality. $T_{mde}$ was at 83% and $T_{max}$ was at 95%. Even at $T_{max}$, nearly 3 functions were still performing well, and nearly 4 at $T_{mde}$. Multiple functions were simultaneously driven to high levels of performance by increasing plant richness.

As shown in the above examples, by putting these metrics together, with the figures showing the relationships between diversity and number of functions above thresholds (Fig. 8), we gain a full picture of how diversity is driving multifunctionality: it is weak in Portugal, moderate in Sweden and Germany, and strong in Sheffield.

*Strengths and weaknesses of the multiple threshold approach*

The suite of metrics generated by the multiple threshold approach provide powerful information for analyzing multifunctionality, especially when combined with analyses of the relationship between diversity and single functions. Reporting these metrics for other studies should also prove useful for future meta-analyses that will allow a quantitative evaluation of the nature and extent of multifunctionality in ecosystems. The multiple threshold approach provides more information and flexibility than any other approach we have reviewed. Overall, the relationship between threshold and the influence of diversity on the number of functions above that threshold provides the fingerprint of diversity's influence on multifunctionality. The multiple threshold approach provides a nuanced view of multifunctionality that allows for direct comparison among experiments and among treatments within an experiment.

The two significant weaknesses we see with this approach are that it provides 1) a suite of metrics rather than a single one and 2) phenomenological rather than mechanistic information. To answer the first, we have concluded that the inherently complex relationships between changing biodiversity and ecosystem functioning are difficult to capture in a single metric. To answer the second, any analysis of multifunctionality must be coupled with analysis of diversity's impact on single functions – or even the impact of key species - to fully understand the mechanisms underlying the observed patterns of diversity effects on multifunctionality.

## CONCLUSIONS

Understanding how changing biodiversity influences the broad suite of processes that ecosystems perform is not simple. Here we compared the most commonly used approaches to characterize multifunctionality. Our analysis shows that systematically exploring how diversity affects multiple functions across the full range of possible thresholds provides an informative "fingerprint" of diversity effects on multifunctionality. The multiple threshold approach provides the most complete and unambiguous summary of the relationships between biodiversity and multifunctionality to date. It addresses many of the ambiguities and problems of previous methods.

Our analysis has focused on how to summarize information on the effects of species richness on multiple ecosystem processes most efficiently and accurately. But understanding multifunctionality mechanistically still requires that such analyses of multifunctionality be complemented with analysis of the effects of species richness on individual functions. The approaches presented here are not the only available analytic tools. Although beyond the scope of our discussion, other approaches such as Structural Equation Modeling (Grace et al. 2010) are potentially promising for incorporating tradeoffs, feedbacks, and other interactions among functions in a more explicit mechanistic manner. Similarly, extrapolating statistical estimates of individual species effects and interactions to simulate and explore untested species compositions may be useful in more thoroughly investigating effects of diversity on multifunctionality. This latter approach can be particularly promising in the presence of complex nonlinearities and species interactions.

The field of biodiversity and ecosystem multifunctionality is still relatively data poor compared to explorations of biodiversity effects on single ecosystem functions. In no small part, this is due to the complex issues generated by the analysis of multifunctionality, the effort to conduct experiments with many levels of species richness, and the difficulty of measuring more than a handful of functions. These logistical issues are surmountable. What is important, now, is to use a common analytical framework to better enable comparisons among experiments as more information becomes available. With results of our comparative analysis in hand, we hope that use of the tools and techniques outlined here and implemented in the *multifunc* package for R will assist in amassing a solid body of data, amenable to investigation of overall trends and underlying mechanisms. We look forward to seeing the field advance.

**Acknowledgements**

This work resulted from the NCEAS working group "Biodiversity and the Functioning of Ecosystems: Translating Results from Model Experiments into Functional Reality". Support for NCEAS comes from University of California Santa Barbara and the National Science Foundation. We thank three anonymous reviewers for their comments. P. Balvanera, K Matulich, and A. Paquette provided valuable feedback on earlier drafts. J.E.K.B. was supported by a postdoctoral fellowship at NCEAS. J.E.D. had support from NSF OCE-1031061; B.J.C. had support from NSF DEB-1046121; L.G. was supported by grant 621-2009-5457 from the Swedish Research Council VR. A Hector was supported by Microsoft Research Cambridge UK. L. Dee was supported by a NSF GRFP DGE-1144085.

## References

Aarssen, L.W. (1997). High productivity in grassland ecosystems: effected by species diversity or productive species? *Oikos*, **80**, 183–184.

Balvanera, P., Pfisterer, A.B., Buchmann, N., He, J.-S., Nakashizuka, T., Raffaelli, D. & Schmid, B. (2006). Quantifying the evidence for biodiversity effects on ecosystem functioning and services. *Ecology Letters*, **9**, 1146–1156.

Cardinale, B.J., Matulich, K.L., Hooper, D.U., Byrnes, J.E., Duffy, E., Gamfeldt, L., Balvanera, P., O'Connor, M.I. & Gonzalez, A. (2011). The functional role of producer diversity in ecosystems. *American Journal of Botany*, **98**, 572–592.

Cardinale, B.J., Srivastava, D.S., Duffy, J.E., Wright, J.P., Downing, A.L., Sankaran, M. & Jouseau, C. (2006). Effects of biodiversity on the functioning of trophic groups and ecosystems. *Nature*, **443**, 989–992.

Duffy, J.E., Richardson, J.P. & Canuel, E.A. (2003). Grazer diversity effects on ecosystem functioning in seagrass beds. *Ecology letters*, **6**, 637–645.

Gamfeldt, L. & Hillebrand, H. (2008). Biodiversity Effects on Aquatic Ecosystem Functioning - Maturation of a New Paradigm. *International Review of Hydrobiology*, **93**, 550–564.

Gamfeldt, L., Hillebrand, H. & Jonsson, P.R. (2008). Multiple functions increase the importance of biodiversity for overall ecosystem functioning. *Ecology*, **89**, 1223–1231.

Grace, J.B., Anderson, T.M., Olff, H. & Scheiner, S.M. (2010). On the specification of structural equation models for ecological systems. *Ecological Monographs*, **80**, 67–87.

He, J.Z., Ge, Y., Xu, Z. & Chen, C. (2009). Linking soil bacterial diversity to ecosystem multifunctionality using backward-elimination boosted trees analysis. *Journal of Soils and Sediments*, **9**, 547–554.

Hector, A. & Bagchi, R. (2007). Biodiversity and ecosystem multifunctionality. *Nature*, **448**, 188–190.

Hooper, D.U. & Vitousek, P.M. (1998). Effects of plant composition and diversity on nutrient cycling. *Ecological Monographs*, **68**, 121–149.

Hooper, D.U., Adair, E.C., Cardinale, B.J., Byrnes, J.E.K., Hungate, B.A., Matulich, K.L., Gonzalez, A., Duffy, J.E., Gamfeldt, L. & O'Connor, M.I. (2012). A global synthesis reveals biodiversity loss as a major driver of ecosystem change. *Nature*, **486**, 105–108.

Hooper, D.U., Chapin, F.S.I., Ewel, J.J., Hector, A., Inchausti, P., Lavorel, S.,

Lawton, J.H., Lodge, D.M., Loreau, M., Naeem, S., Schmid, B., Setälä, H., Symstad, A.J., Vandermeer, J. & Wardle, D.A. (2005). Effects of biodiversity on ecosystem functioning: a consensus of current knowledge. *Ecological Monographs*, **75**, 3–35.

Isbell, F., Calcagno, V., Hector, A., Connolly, J., Harpole, W.S., Reich, P.B., Scherer-Lorenzen, M., Schmid, B., Tilman, D., van Ruijven, J., Weigelt, A., Wilsey, B.J., Zavaleta, E.S. & Loreau, M. (2011). High plant diversity is needed to maintain ecosystem services. *Nature*, **477**, 199–202.

Loreau, M. & Hector, A. (2001). Partitioning selection and complementarity in biodiversity experiments. *Nature*, **412**, 72–76.

Maestre, F.T., Castillo-Monroy, A.P., Bowker, M.A. & Ochoa-Hueso, R. (2012a). Species richness effects on ecosystem multifunctionality depend on evenness, composition and spatial pattern. *Journal of Ecology*, **100**, 317–330.

Maestre, F.T., Quero, J.L., Gotelli, N.J., Escudero, A., Ochoa, V., Delgado-Baquerizo, M., Garcia-Gomez, M., Bowker, M.A., Soliveres, S., Escolar, C., Garcia-Palacios, P., Berdugo, M., Valencia, E., Gozalo, B., Gallardo, A., Aguilera, L., Arredondo, T., Blones, J., Boeken, B., Bran, D., Conceicao, A.A., Cabrera, O., Chaieb, M., Derak, M., Eldridge, D.J., Espinosa, C.I., Florentino, A., Gaitan, J., Gatica, M.G., Ghiloufi, W., Gomez-Gonzalez, S., Gutierrez, J.R., Hernandez, R.M., Huang, X., Huber-Sannwald, E., Jankju, M., Miriti, M., Monerris, J., Mau, R.L., Morici, E., Naseri, K., Ospina, A., Polo, V., Prina, A., Pucheta, E., Ramirez-Collantes, D.A., Romao, R., Tighe, M., Torres-Diaz, C., Val, J., Veiga, J.P., Wang, D. & Zaady, E. (2012b). Plant Species Richness and Ecosystem Multifunctionality in Global Drylands. *Science*, **335**, 214–218.

Mouillot, D., Villeger, S., Scherer-Lorenzen, M. & Mason, N.W.H. (2011). Functional Structure of Biological Communities Predicts Ecosystem Multifunctionality. *PLoS one*, **6**, e17476.

O'Hara, R.B. & Kotze, D.J. (2010). Do not log-transform count data. *Methods in Ecology and Evolution*, **1**, 118–122.

Peter, H., Ylla, I., Gudasz, C., Romaní, A.M., Sabater, S. & Tranvik, L.J. (2011). Multifunctionality and Diversity in Bacterial Biofilms. *PLoS one*, **6**, e23225.

Reich, P.B., Tilman, D., Isbell, F., Mueller, K., Hobbie, S.E., Flynn, D.F.B. & Eisenhauer, N. (2012). Impacts of Biodiversity Loss Escalate Through Time as Redundancy Fades. *Science*, **336**, 589–592.

Scherer-Lorenzen, M., Palmborg, C., Prinz, A. & Schulze, E.-D. (2003). The role of plant diversity and composition for nitrate leaching in grasslands. *Ecology*, **84**, 1539–1552.

Spehn, E.M., Hector, A., Joshi, J., Scherer-Lorenzen, M., Schmid, B., Bazeley-White, E., Beierkuhnlein, C., Caldeira, M.C., Diemer, M., Dimitrakopoulos, P.G., Finn, J.A., Freitas, H., Giller, P.S., Good, J., Harris, R., Högberg, P., Huss-Danell, K., Jumpponen, A., Koricheva, J., Leadley, P.W., Loreau, M., Minns, A., Mulder, C.P.H., O'Donovan, G., Otway, S.J., Palmborg, C., Pereira, J.S., Pfisterer, A.B., Prinz, A., Read, D.J., Schulze, E.D., Siamantziouras, A.S., Terry, A.C., Troumbis, A.Y., Woodward, F.I., Yachi, S. & Lawton, J.H. (2005). Ecosystem effects of biodiversity manipulations in European grasslands. *Ecological Monographs*, **75**, 37–63.

Stachowicz, J.J., Bruno, J.F. & Duffy, J.E. (2007). Understanding the Effects of Marine Biodiversity on Communities and Ecosystems. *Annual Review of Ecology Evolution and Systematics*, **38**, 739–766.

Tilman, D., Reich, P.B. & Isbell, F. (2012). Biodiversity impacts ecosystem productivity as much as resources, disturbance, or herbivory. *Proceedings of the National Academy of Sciences of the United States of America*, 10394–10397.

Ver Hoef, J.M. & Boveng, P.L. (2007). Quasi-poisson vs. negative binomial regression: how should we model overdispersed count data? *Ecology*, **88**, 2766–2772.

Warton, D. (2011). The arcsine is asinine: the analysis of proportions in ecology. *Ecology*.

Zavaleta, E.S., Pasari, J.R., Hulvey, K.B. & Tilman, G.D. (2010). Sustaining multiple ecosystem functions in grassland communities requires higher biodiversity. *Proceedings of the National Academy of Sciences*, **107**, 1443–1446.

**Supplementary Information 1:** Code and data used for all analyses and figures in this paper.

**Supplementary Information 2:** Choice of underlying model for threshold approach

**Table 1.** Comparison of four approaches previously used to quantify ecosystem multifunctionality, and the new approach recommended here. The table summarizes what questions are addressed by each approach, what unique information is gained, what the limitations are, and references that have used the approach. For each question in the column "Question addressed", an answer of 'no' would correspond to the null hypothesis, and an answer of 'yes' would correspond to a testable alternative hypothesis

| | Approach | Question addressed | Unique information | Limitations | References |
|---|---|---|---|---|---|
| *Previous approaches* | 1. Single Functions | Do more functions achieve high values in a diverse mixture than for any single species? | • Direct information about each individual function | • Qualitative<br>• Does not provide a metric relating diversity and multifunctionality | Duffy et al. 2003 |
| | 2. Turnover | Do different species promote different functions? | • Indicates whether different species drive different processes | • Does not consider negative effects<br>• Does not measure multifunctionality directly<br>• Requires extensive data | Hector & Bagchi 2007, He et al 2009, Isbell et al 2011 |
| | 3. Averaging | Does the average level of multiple functions increase with the number of species? | • Indicates average diversity effect on functions | • Single functions can have large impact.<br>• Cannot distinguish between (i) two functions at similar level and (ii) one function at high level and other function at low | Hooper & Vitousek 1998, Mouillot et al. 2011, Maestre et al. 2012a,b |
| | | | | level | |
| | 4. Single Threshold | Does the number of functions exceeding a threshold increase with the number of species? | • Indicates whether multiple functions have high value | • Threshold is arbitrary<br>• Does not indicate extent to which threshold is exceeded or not | Gamfeldt et al 2008, Zavaleta et al 2010, Peter et al. 2011 |
| *Our Approach* | 5. Multiple Thresholds | Does diversity influence the level of performance of multiple functions? | • Provides a measure of how diversity simultaneously influences multiple functions<br>• Multiple informative metrics describe different aspects of multifunctionality | • Produces a curve rather than a single number | This paper |

**Table 2.** Values for indices generated by multiple threshold approach to multifunctionality from analyses for Germany, Portugal, Sweden, and Sheffield. The characters -- indicates a value that could not be calculated (e.g., there is no maximum value where the relationship between diversity and number of functions again becomes 0). Definitions of indices are in the text.

| Location | $T_{min}$ | $T_{max}$ | $T_{mde}$ | $R_{mde}$ | $P_{mde}$ | $M_{min}$ | $M_{max}$ | $M_{mde}$ |
|---|---|---|---|---|---|---|---|---|
| Germany | 15% | 97% | 50% | 0.16 | 52.79% | 5.06 | 1.14 | 4.47 |
| Portugal | 9% | 49% | 32% | 0.22 | 50.80% | 6.10 | 4.11 | 6.44 |
| Sweden | -- | 73% | 46% | 0.27 | 53.53% | -- | 2.20 | 4.78 |
| Sheffield | 58% | 95% | 83% | 0.19 | 58.39% | 4.26 | 2.71 | 3.89 |

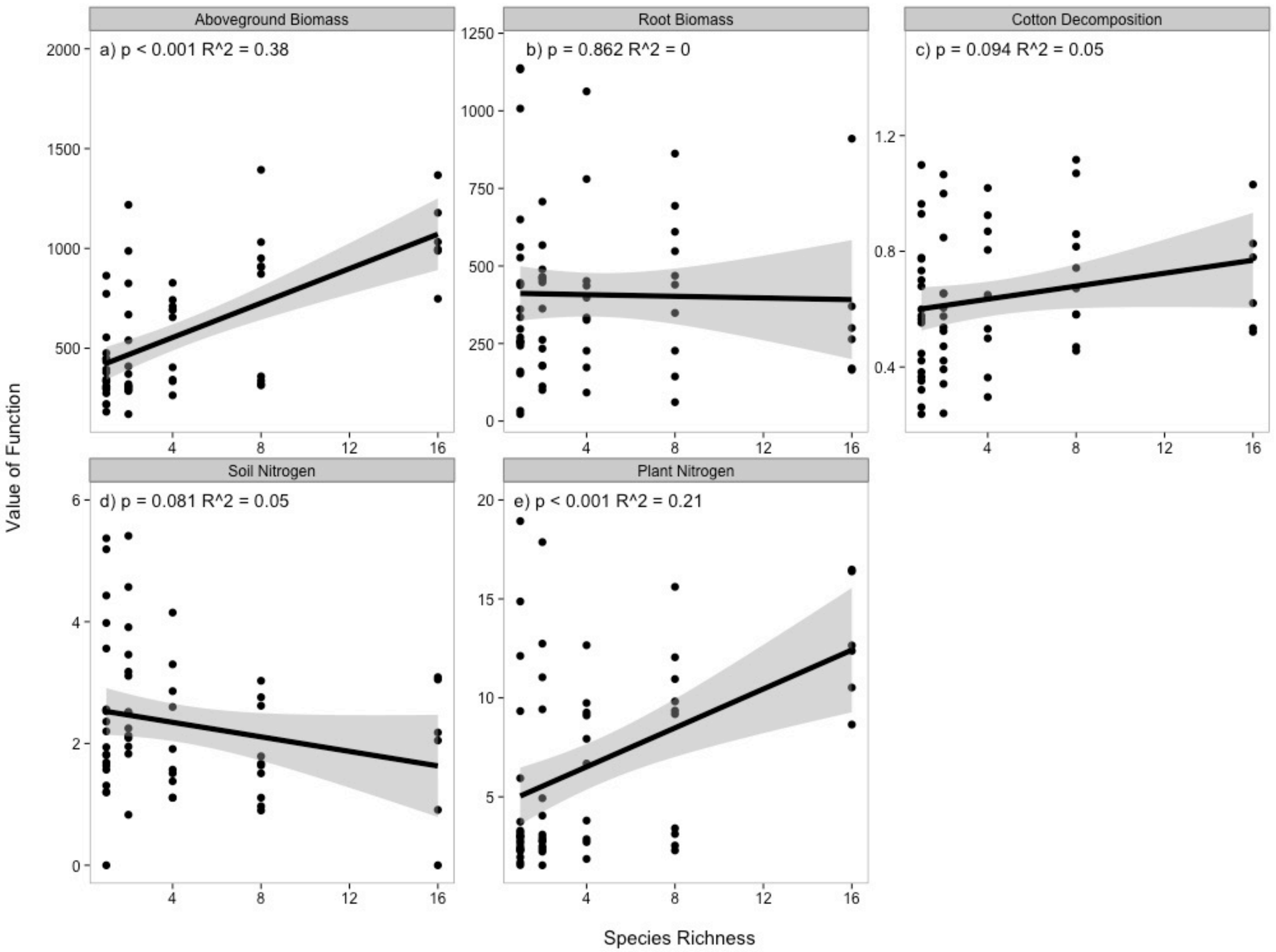

Fig 1: The relationship between species richness and the values of functions measured at the BIODEPTH Germany site (a-e). Note that original analyses were on a log scale, but the qualitative results do not differ.

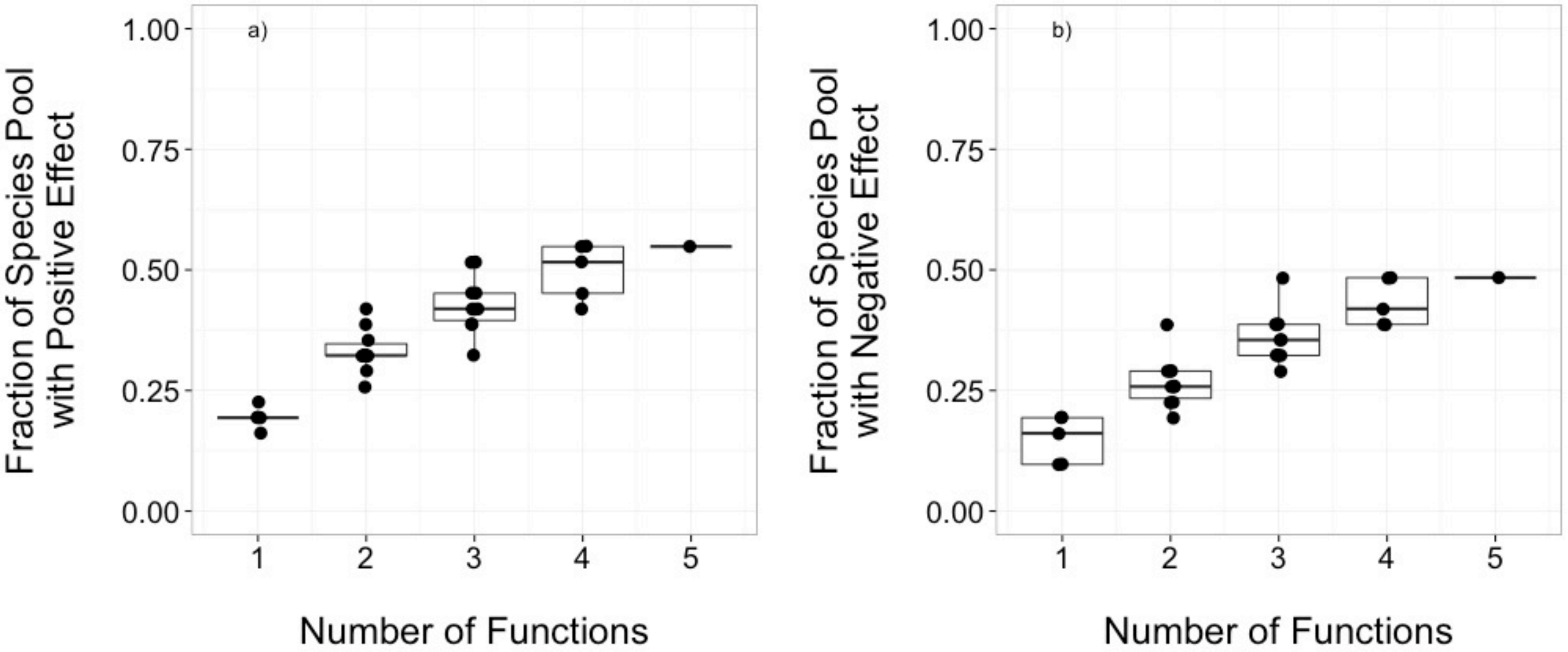

Fig 2: The relationship between the number of functions considered and the fraction of the species pool having either (a) positive or (b) negative effects on ecosystem functions. Points for each combination of functions are overlaid on top of a box and whisker plot providing medians and interquartile ranges . Points have been jittered in both dimensions so that combinations with overlapping values can be seen.

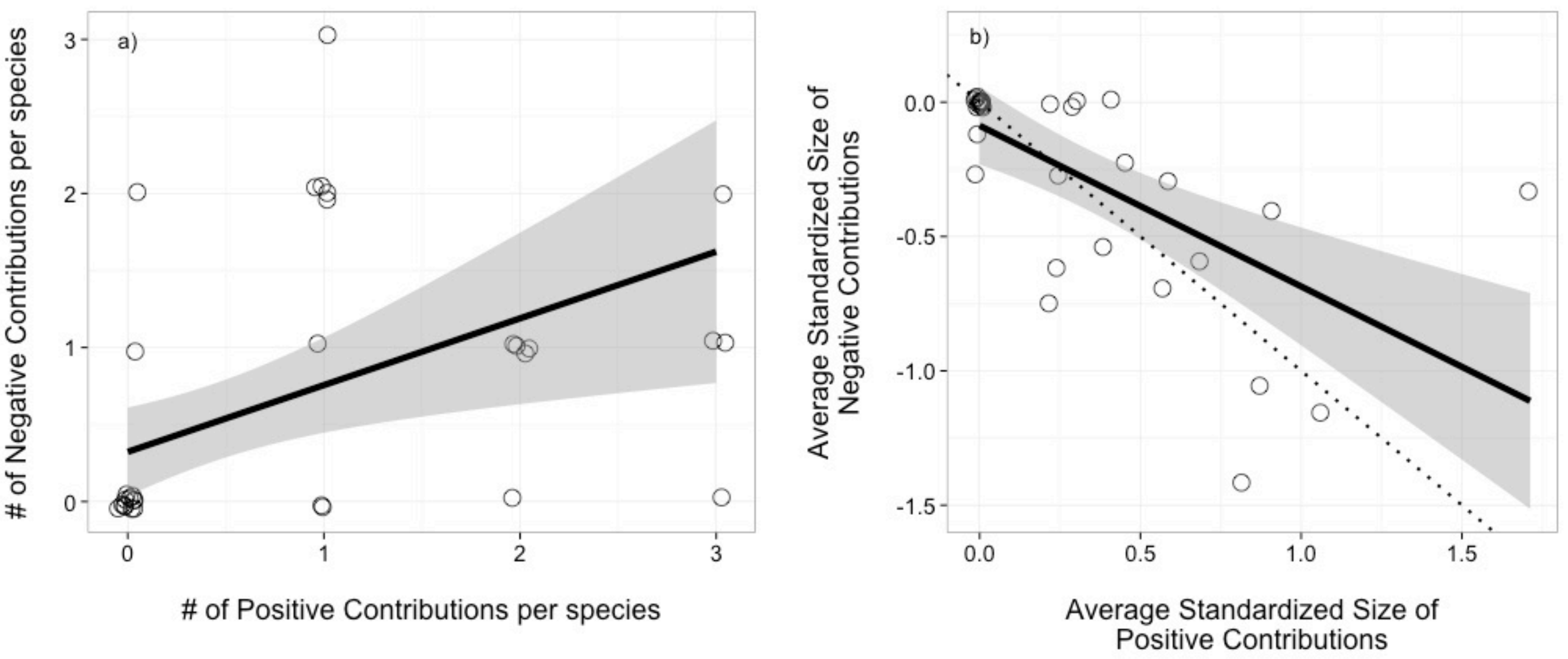

Fig 3: The relative balance of positive and negative contributions of species. Each point is a species. Note, all points have been jittered in order to see overlapping data points. The relationship between a) total number of positive and negative contributions of species to different ecosystem functions. Panel b) shows the relationship between the average positive effect and the average negative effect of each species. The dashed line is a -1:1 relationship. Shaded areas show the 95% CI of the fit.

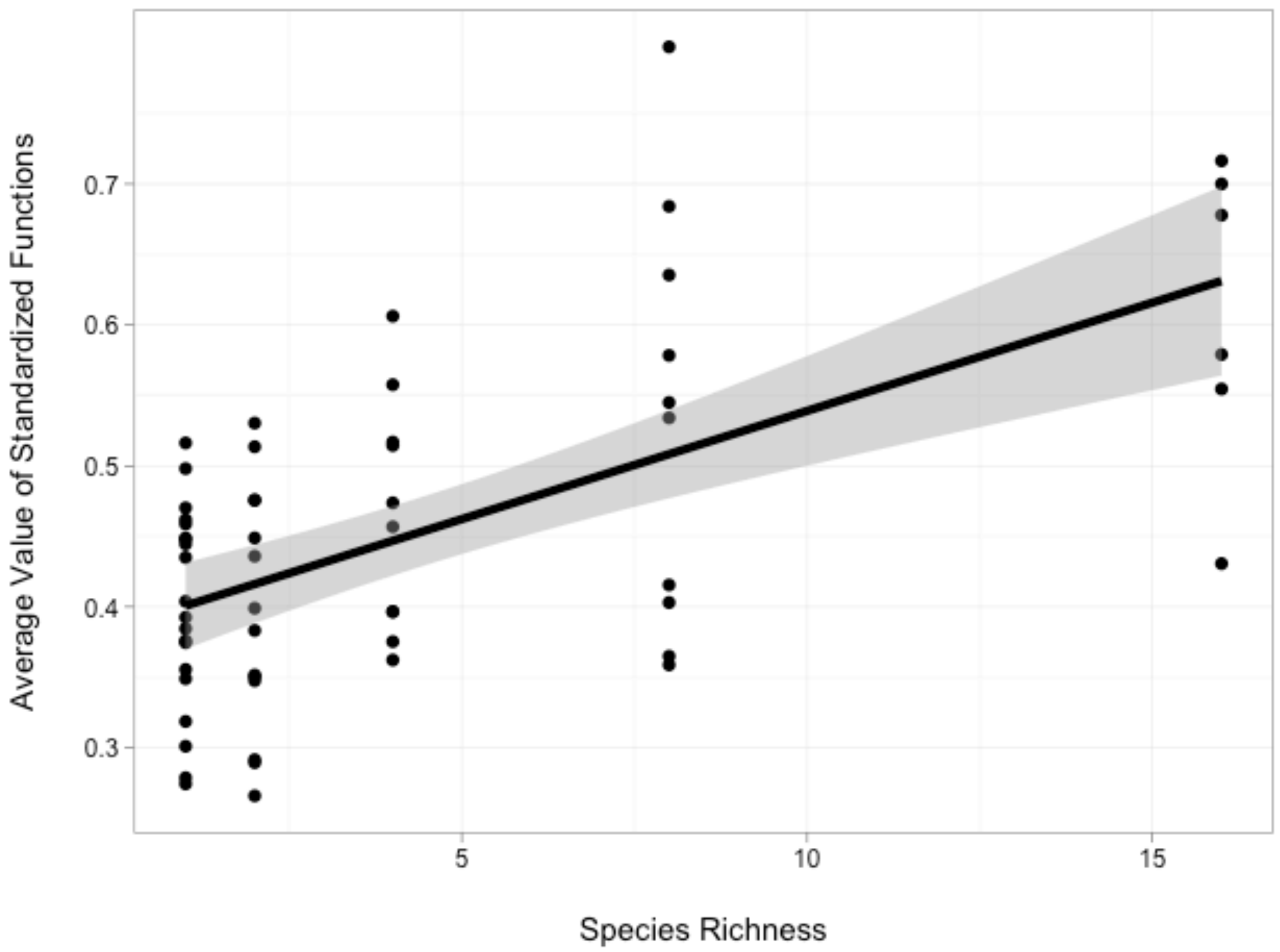

Fig 4: The relationship between the species richness and the average of the standardized value of functions ($MF_a$) measured at the BIODEPTH Germany. Slope estimate = 0.0133 ± 0.0028 SE , Diversity $F_{1,22}$=22.554, $p<0.001$, $R^2$=0.36.

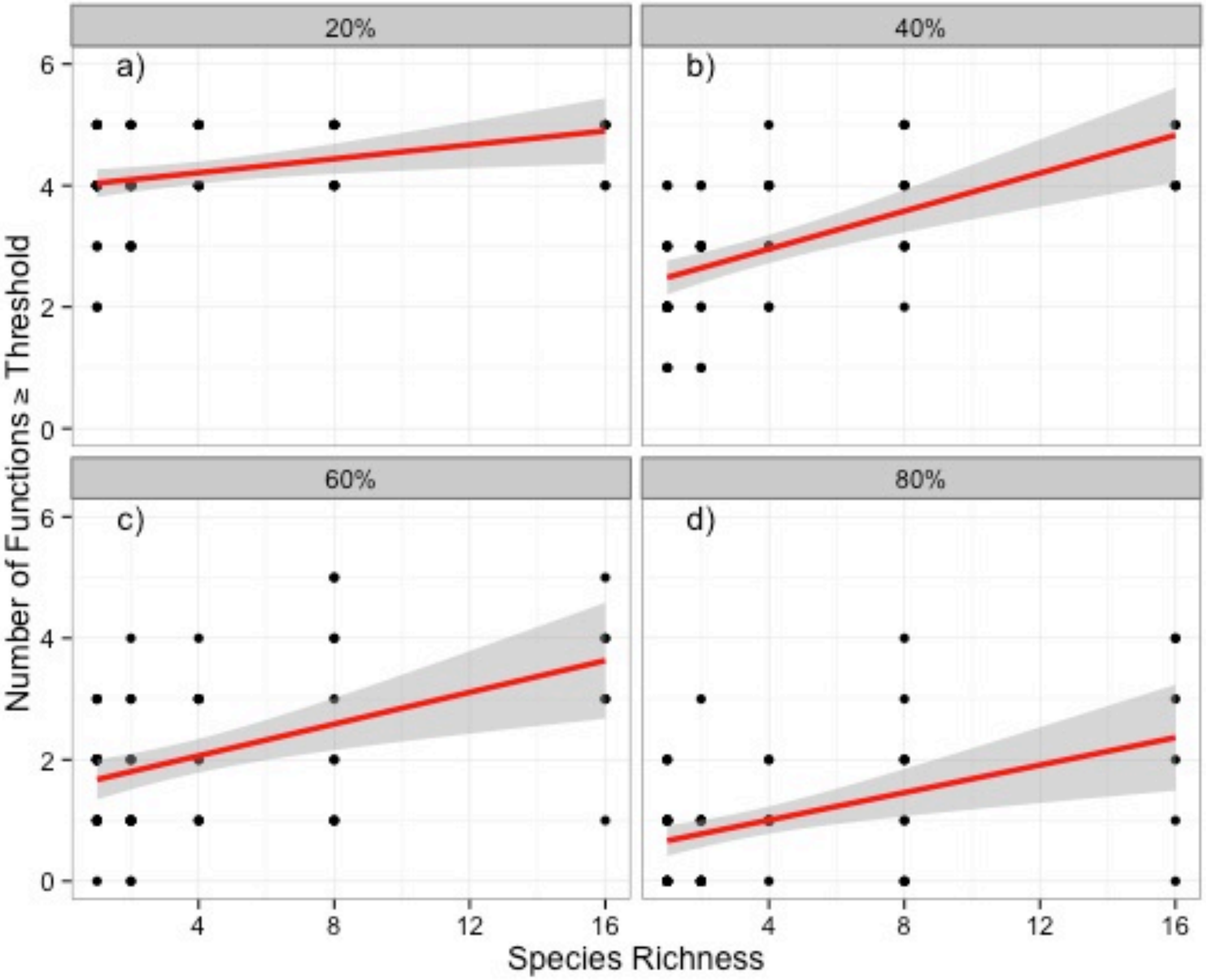

Fig 5: The relationship between planted species richness and multifunctionality, defined as number of function reaching a threshold of some proportion of the maximum observed function.  Panels show the relationship for four different thresholds (0.2, 0.4, 0.6, and 0.8 of maximum) in plots in the German portion of the BIODEPTH experiment.

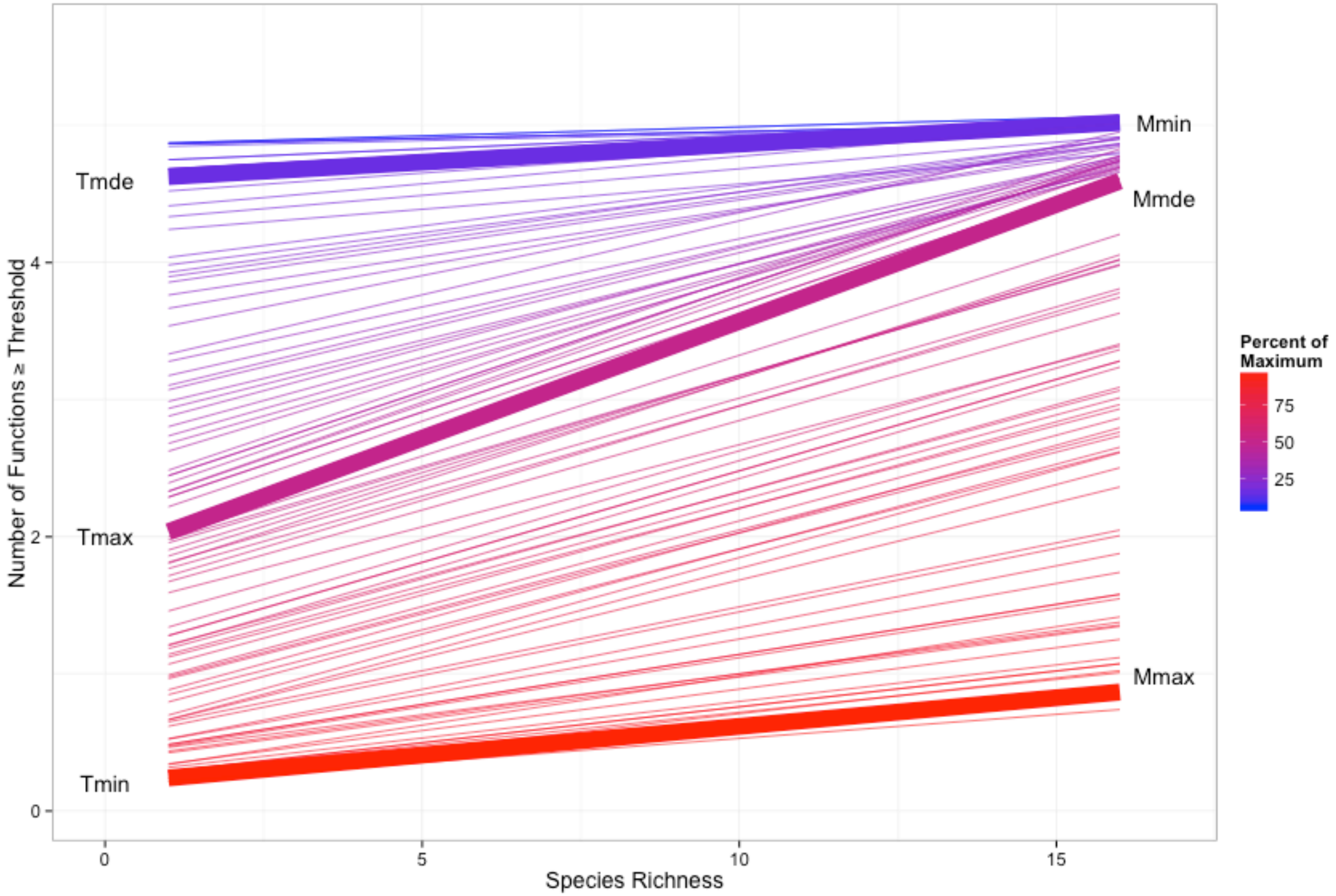

Fig 6: The relationship between planted species richness and the number of functions at or above a threshold of some proportion of the maximum observed function. Colors indicate different thresholds as shown in the figure legend with cooler colors denoting lower thresholds and warmer colors denoting higher thresholds. Data are from the German portion of the BIODEPTH experiment. Tmin is the line with the lowest threshold whose slope is different from 0. Tmde is the line with the steepest slope. Tmax is the line at the highest threshold where the slope is different from 0. All indices preceded by M indicate the number of functions for the corresponding curves.

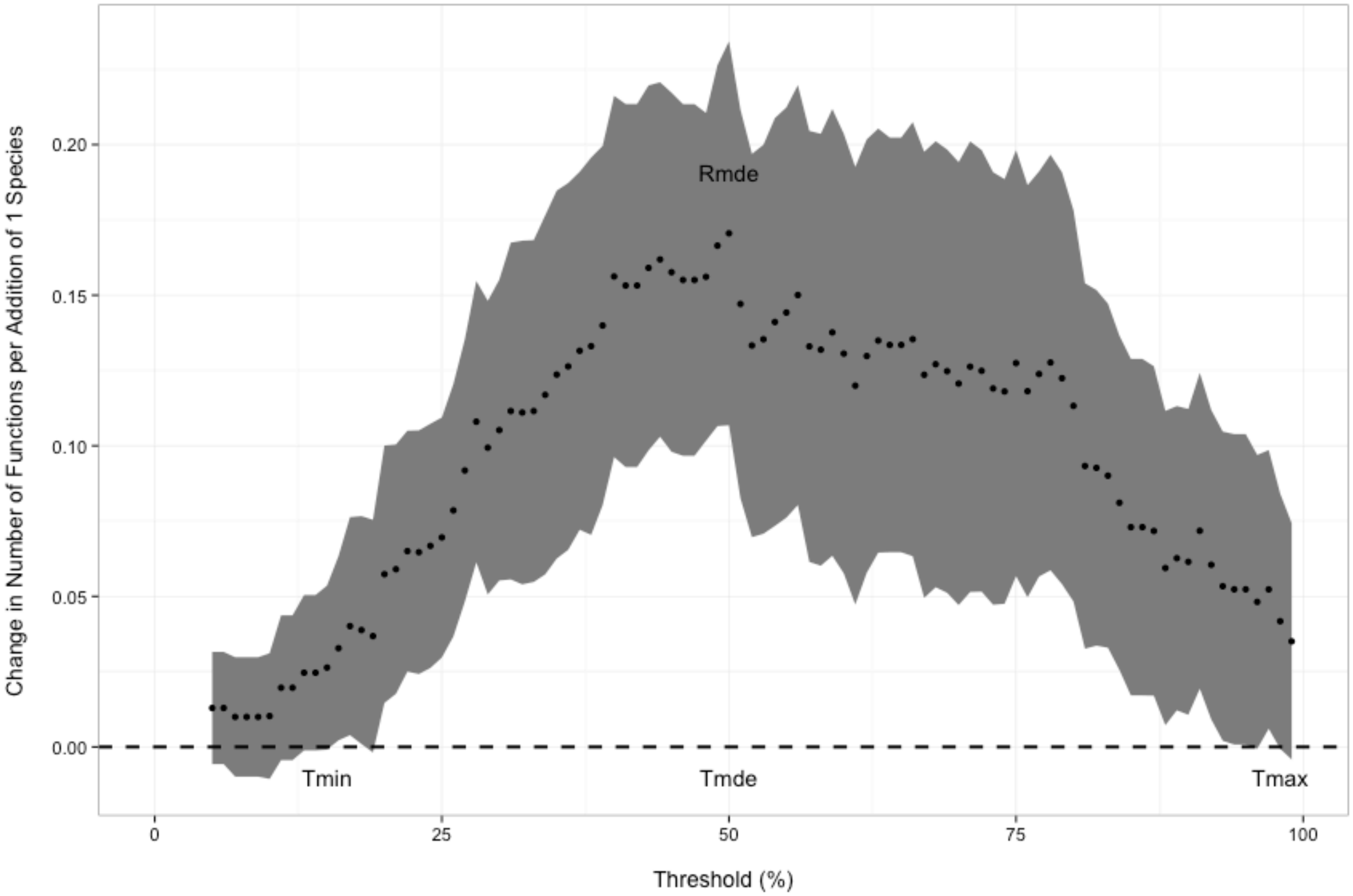

Fig 7: The slope of the relationship between planted species richness and the number of functions at or above a threshold of some proportion of the maximum observed function, at different threshold values (x-axis). Points are the fitted values and shading indicated +/- 1 CI. Data are from the German portion of the BIODEPTH experiment. Tmin is the slope with the lowest threshold that is not from 0. Tmde is the threshold with the steepest slope. Tmaxis the maximum threshold where the slope again becomes no different from 0. Rmde shows the maximum slope estimated at Tmde

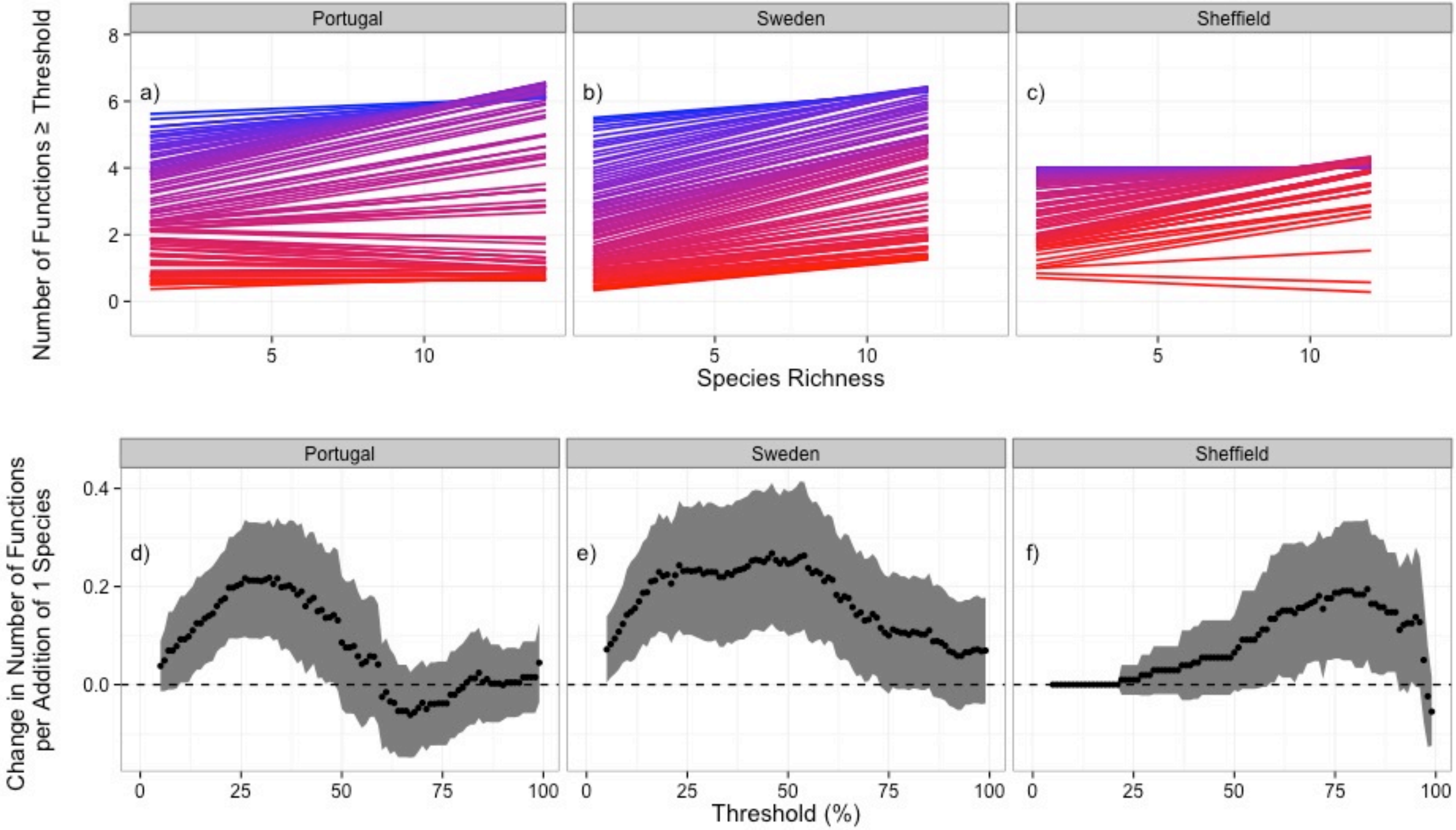

Fig 8: Multifunctionality analyses at multiple BIODEPTH sites. Panels a-c show the relationships between planted species richness and the number of functions above a threshold for multiple different threshold values. Colors indicate different thresholds as shown in the legend for Fig. 6 (blue=low, red=high) with cooler colors denoting lower thresholds and warmer colors denoting higher thresholds. Panels d-f show the corresponding relationship between threshold value and slope of the relationship between planted species richness and the number of functions reaching a threshold. Points are the fitted values and shading indicates +/- 1 CI.

# Supplementary Information 1: Using the *multifunc* package for analysis of Biodiversity Ecosystem Multifunctionality Relationships in R

This is a step-by-step walk-through of the analysis of the German BIODEPTH site in the paper to illustrate the different analyses we conducted to evaluate the relationship between biodiversity and ecosystem multifunctionality.

We begin by loading the *multifunc* package. If it is not currently installed, it can be found on github, and installed using the *devtools* library.

```
library(devtools)
install_github("multifunc", username="jebyrnes")
```

```
library(multifunc)

#for plotting
library(ggplot2)
library(gridExtra)
```

Next, we load in the BIODEPTH data. We use some of the helper functions in the package to identify the column numbers of the relevant functions for later use.

```
#Read in data  to run sample analyses on the biodepth data
data(all_biodepth)

allVars<-qw(biomassY3, root3, N.g.m2,  light3, N.Soil, wood3, cotton3)
varIdx<-which(names(all_biodepth) %in% allVars)
```

As we want to work with just the German data, we next subset down the data to just Germany. We then look at the whether species were seeded or not, and determine which species were indeed seeded in Germany. If a column of species is full of zeroes, then it was not seeded in these plots. We will need the column indices of relevant species for later use in the overlap analyses.

```
#######
#Now, specify what data we're working with - Germany
#and the relevant variables we'll be working with
#######
germany<-subset(all_biodepth, all_biodepth$location=="Germany")


vars<-whichVars(germany, allVars)
species<-relevantSp(germany,26:ncol(germany))
spIDX <- which(names(germany) %in% species) #in case we need these
```

**Single Function Approach**

First, we will demonstrate the qualitative single function approach. As we're using *ggplot2* for plotting, we'll use *reshape* to melt the data into something suitable to make a nice faceted plot of each relationship.

```
germanyForPlotting<-melt(germany[,c(8,which(names(germany) %in% vars))],
id.vars="Diversity")
germanyForPlotting$variable <- factor(germanyForPlotting$variable)

#make the levels of the functions into something nice for plots
levels(germanyForPlotting$variable) <- c('Aboveground Biomass', 'Root
Biomass', 'Cotton Decomposition', 'Soil Nitrogen', 'Plant Nitrogen')
```

Nest, as we want to display additional information about the fit of diversity to each function, we'll need to iterate over each function, derive a fit, and then make some labels to be used in the eventual plot.

```
germanyFits <- dlply(germanyForPlotting, .(variable), function(x) lm(value ~
Diversity, data=x))
germanyLabels <- data.frame(variable = levels(germanyForPlotting$variable),
                            Diversity=7, value=c(2000, 1200, 1.5, 6, 20),
                            lab=paste(letters[1:5], ")", sep=""),
                            r2 = sapply(germanyFits, function(x)
summary(x)$r.squared),
                            p = sapply(germanyFits, function(x)
anova(x)[1,5])
                            )

germanyLabels$labels <- with(germanyLabels, paste(lab, "p =",
round(p,3),expression(R^2), "=", round(r2,2),   sep=" "))
germanyLabels$labels <- gsub("p = 0 ", "p < 0.001 ", germanyLabels$labels)
```

Finally, we will put it all together into a single plot.

```
ggplot(aes(x=Diversity, y=value),data=germanyForPlotting) +
  geom_point(size=3)+
  facet_wrap(~variable, scales="free") +
  theme_bw(base_size=15)+
  stat_smooth(method="lm", colour="black", size=2) +
  xlab("\nSpecies Richness") +
  ylab("Value of Function\n") +
  geom_text(data=germanyLabels,
            aes(label=labels)) +
 theme(panel.grid = element_blank())
```

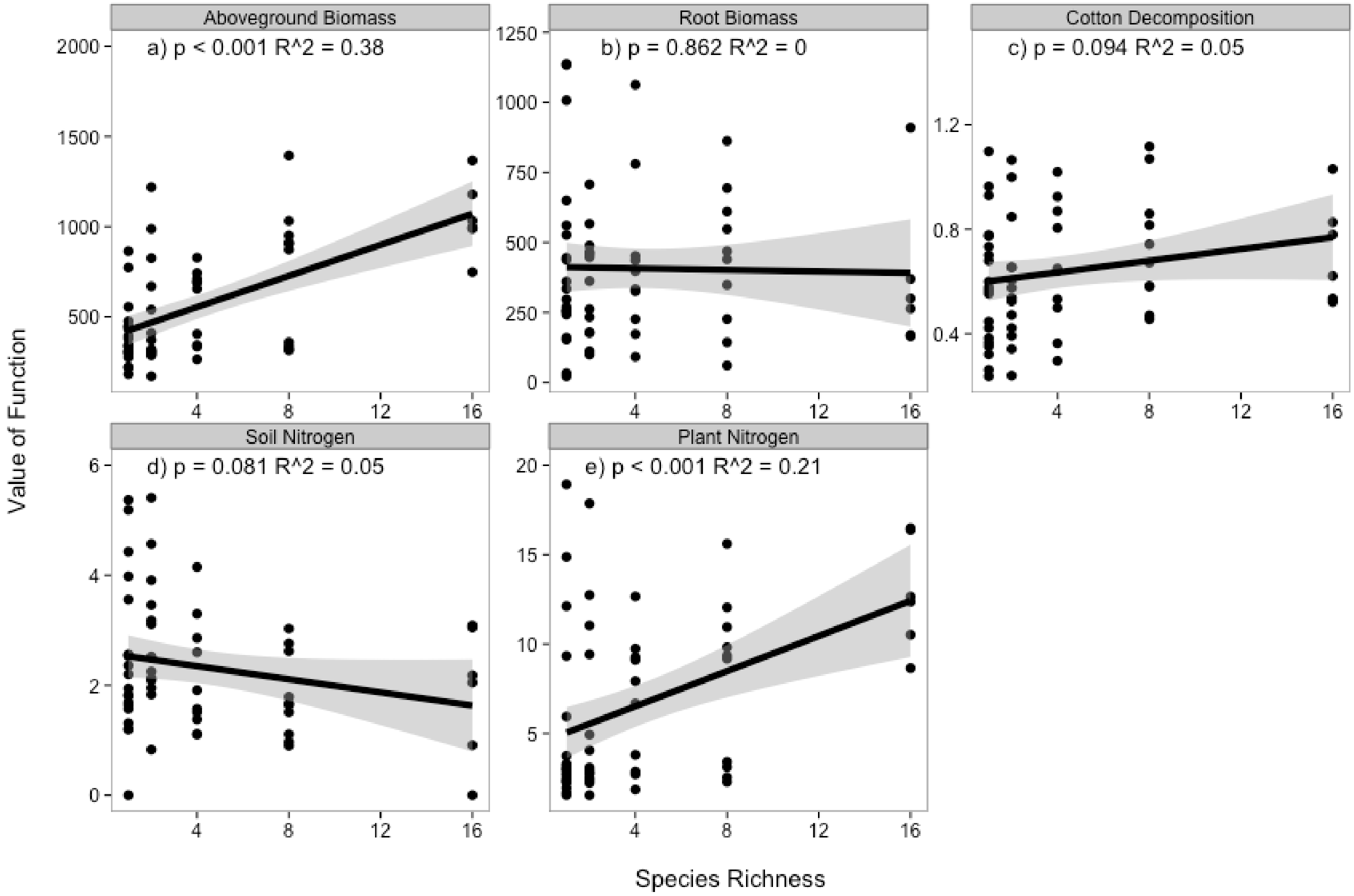

Next, we ask whether the best performing species differs between functions. To do that, we have to pull out all monoculture plots, then reshape the data and run a quick qualitative analysis on it to determine the best performer.

```
#pull out only those plots planted in monoculture
monoGermany<- subset(germany, rowSums(germany[,spIDX])==1)

#figure out which monoculture is which
monoGermany$mono<-apply(monoGermany[,spIDX], 1, function(x)
species[which(x==1)])

#melt by function for easier analysis
monoGermanyMelt<-melt(monoGermany[,c(ncol(monoGermany),which(names(monoGermany)
 %in% vars))], id.vars="mono")

#get the mean for each monoculture that has >1 replicate
monoGermanyMelt <- ddply(monoGermanyMelt, .(variable, mono), summarize,
value=mean(value))

#now, who is the best performer?
ddply(monoGermanyMelt, .(variable), summarize,
maxMono=mono[which(value==max(value, na.rm=T))])
```

```
#    variable maxMono
# 1 biomassY3 TRIPRA1
# 2     root3 FESRUB1
# 3   cotton3 TRIREP1
# 4    N.Soil TRIREP1
# 5    N.g.m2 TRIPRA1
```

**Overlap Approach**

One of the functions we are looking at, soil N, first needs to be reflected before we can apply the overlap approach.

```
germany$N.Soil<- -1*germany$N.Soil +max(germany$N.Soil, na.rm=T)
```

We can examine any single species and generate coefficient values and 'effects' - i.e., whether a species has a positive (1), negative (-1), or neutral (0) effect on a function. Let's examine this for biomass production in year 3.

```
spList<-sAICfun("biomassY3", species, germany)
spList
```

```
# $pos.sp
# [1] "DACGLO1" "FESPRA1" "LATPRA1" "LOTCOR1" "TRIPRA1" "TRIREP1"
#
# $neg.sp
# [1] "CAMPAT1" "PHLPRA1" "RANACR1"
#
# $neu.sp
#  [1] "ACHMIL1" "ALOPRA1" "ANTODO1" "ARRELA1" "BROHOR1" "CENJAC1" "CHRLEU1"
#  [8] "CREBIE1" "CYNCRI1" "FESRUB1" "GERPRA1" "HOLLAN1" "KNAARV1" "LEOAUT1"
# [15] "LOLPER1" "LYCFLO1" "PIMMAJ1" "PLALAN1" "RUMACE1" "TAROFF1" "VICCRA1"
# [22] "VICSEP1"
#
# $functions
# [1] "biomassY3"
#
# $coefs
#     ACHMIL1     ALOPRA1     ANTODO1     ARRELA1     BROHOR1     CAMPAT1
#         0.0         0.0         0.0         0.0         0.0      -391.5
#     CENJAC1     CHRLEU1     CREBIE1     CYNCRI1     DACGLO1     FESPRA1
#         0.0         0.0         0.0         0.0       104.6       135.7
#     FESRUB1     GERPRA1     HOLLAN1     KNAARV1     LATPRA1     LEOAUT1
#         0.0         0.0         0.0         0.0       248.6         0.0
#     LOLPER1     LOTCOR1     LYCFLO1     PHLPRA1     PIMMAJ1     PLALAN1
#         0.0       375.7         0.0      -297.7         0.0         0.0
#     RANACR1     RUMACE1     TAROFF1     TRIPRA1     TRIREP1     VICCRA1
#      -140.3         0.0         0.0       492.6       325.0         0.0
#     VICSEP1 (Intercept)
#         0.0       335.8
#
# $effects
# ACHMIL1 ALOPRA1 ANTODO1 ARRELA1 BROHOR1 CAMPAT1 CENJAC1 CHRLEU1 CREBIE1
#       0       0       0       0       0      -1       0       0       0
# CYNCRI1 DACGLO1 FESPRA1 FESRUB1 GERPRA1 HOLLAN1 KNAARV1 LATPRA1 LEOAUT1
#       0       1       1       0       0       0       0       1       0
# LOLPER1 LOTCOR1 LYCFLO1 PHLPRA1 PIMMAJ1 PLALAN1 RANACR1 RUMACE1 TAROFF1
#       0       1       0      -1       0       0      -1       0       0
# TRIPRA1 TRIREP1 VICCRA1 VICSEP1
#       1       1       0       0
```

Using this as our starting point, we can use the getRedundancy function to generate a) an effect matrix for all functions, b) a coefficient matrix for all functions, and c) a standardized coefficient matrix for all functions.

```
redund<-getRedundancy(vars, species, germany)
coefs<-getRedundancy(vars, species, germany, output="coef")
stdCoefs<-stdEffects(coefs, germany, vars, species)

#for example
redund
```

```
#             ACHMIL1 ALOPRA1 ANTODO1 ARRELA1 BROHOR1 CAMPAT1 CENJAC1 CHRLEU1
# biomassY3         0       0       0       0       0      -1       0       0
# root3             1       0      -1       0       1       0       0       0
# N.g.m2            1       0       0      -1       0       0       0       0
# N.Soil           -1       0      -1       0       1      -1       0       0
# cotton3           0       1       1       0      -1       1       0       0
#             CREBIE1 CYNCRI1 DACGLO1 FESPRA1 FESRUB1 GERPRA1 HOLLAN1 KNAARV1
# biomassY3         0       0       1       1       0       0       0       0
# root3             0       0       0       0       1       0      -1       0
# N.g.m2           -1       0       0       1       0       0      -1       0
# N.Soil            1       0       1      -1       0       0       0       0
# cotton3           0       0       0       0       0       0       0       0
#             LATPRA1 LEOAUT1 LOLPER1 LOTCOR1 LYCFLO1 PHLPRA1 PIMMAJ1 PLALAN1
# biomassY3         1       0       0       1       0      -1       0       0
# root3             1       0       1       0       0       0       0      -1
# N.g.m2            0       0       1       1       0      -1       0      -1
# N.Soil            1       0       0       1       0       1       0       0
# cotton3          -1       0      -1       0       0       0       0       1
#             RANACR1 RUMACE1 TAROFF1 TRIPRA1 TRIREP1 VICCRA1 VICSEP1
# biomassY3        -1       0       0       1       1       0       0
# root3            -1       0       0      -1       0       0       0
# N.g.m2           -1       0       0       1       1       0       0
# N.Soil            1       0       0      -1      -1       0       0
# cotton3           0       0       0       1       1       0       0
```

Using the effect matrix, we then estimate the average number of species affecting all possible combinations of functions - both positive and negative - and then plot the results.

```
#plot the num. functions by fraction of the species pool needed
posCurve<-divNeeded(redund, type="positive")
posCurve$div<-posCurve$div/ncol(redund)
pc<-qplot(nfunc, div, data=posCurve, group=nfunc, geom=c("boxplot"))+
  geom_jitter(size=4, position = position_jitter(height=0.001, width = .04))+
  ylab("Fraction of Species Pool\nwith Positive Effect\n")+
  xlab("\nNumber of Functions")+theme_bw(base_size=24)+ylim(c(0,1))


negCurve<-divNeeded(redund, type="negative")
negCurve$div<-negCurve$div/ncol(redund)
nc<-qplot(nfunc, div, data=negCurve, group=nfunc, geom=c("boxplot"))+
  geom_jitter(size=4, position = position_jitter(height=0.001, width = .04))+
  ylab("Fraction of Species Pool\nwith Negative Effect\n")+
  xlab("\nNumber of Functions")+theme_bw(base_size=24)+ylim(c(0,1))

#combine these into one plot
grid.arrange(pc+annotate("text", x=1, y=1, label="a)"), nc+annotate("text",
x=1, y=1, label="b)"), ncol=2)
```

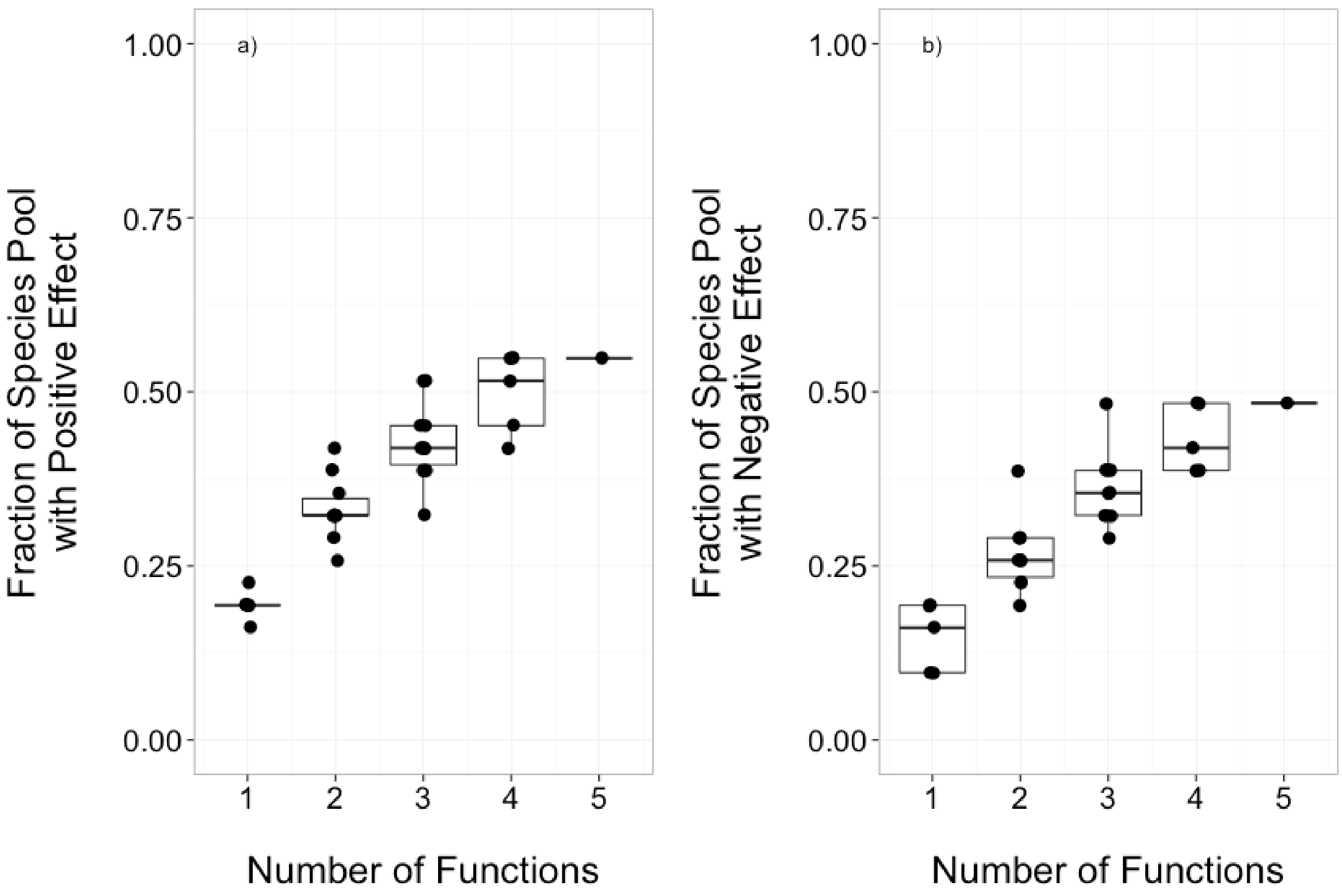

We could have also looked at overlap indices, such as the Sorenson index.

```
allOverlapPos<-ddply(data.frame(m=2:nrow(redund)), .(m), function(adf)
getOverlapSummary(redund, m=adf$m[1], denom="set"))
```

One we've examined the consequences of functional overlap, we can then move on to compare the number and size of these effects. First, what is the relative balance of positive to negative contributions for each species?

```
posNeg<-data.frame(Species = colnames(redund),
                   Pos = colSums(filterOverData(redund)),
                   Neg = colSums(filterOverData(redund, type="negative")))

#plot it
ggplot(aes(x=Pos,y= Neg), data=posNeg) +
  geom_jitter(position = position_jitter(width = 0.05, height = 0.05),
size=5, shape=1) +
  theme_bw(base_size=18) +
  xlab("\n# of Positive Contributions per species\n") +
  ylab("# of Negative Contributions per species\n") +
  annotate("text", x=0, y=3, label="a)") +
  stat_smooth(method="glm", colour="black", size=2,
family=poisson(link="identity"))
```

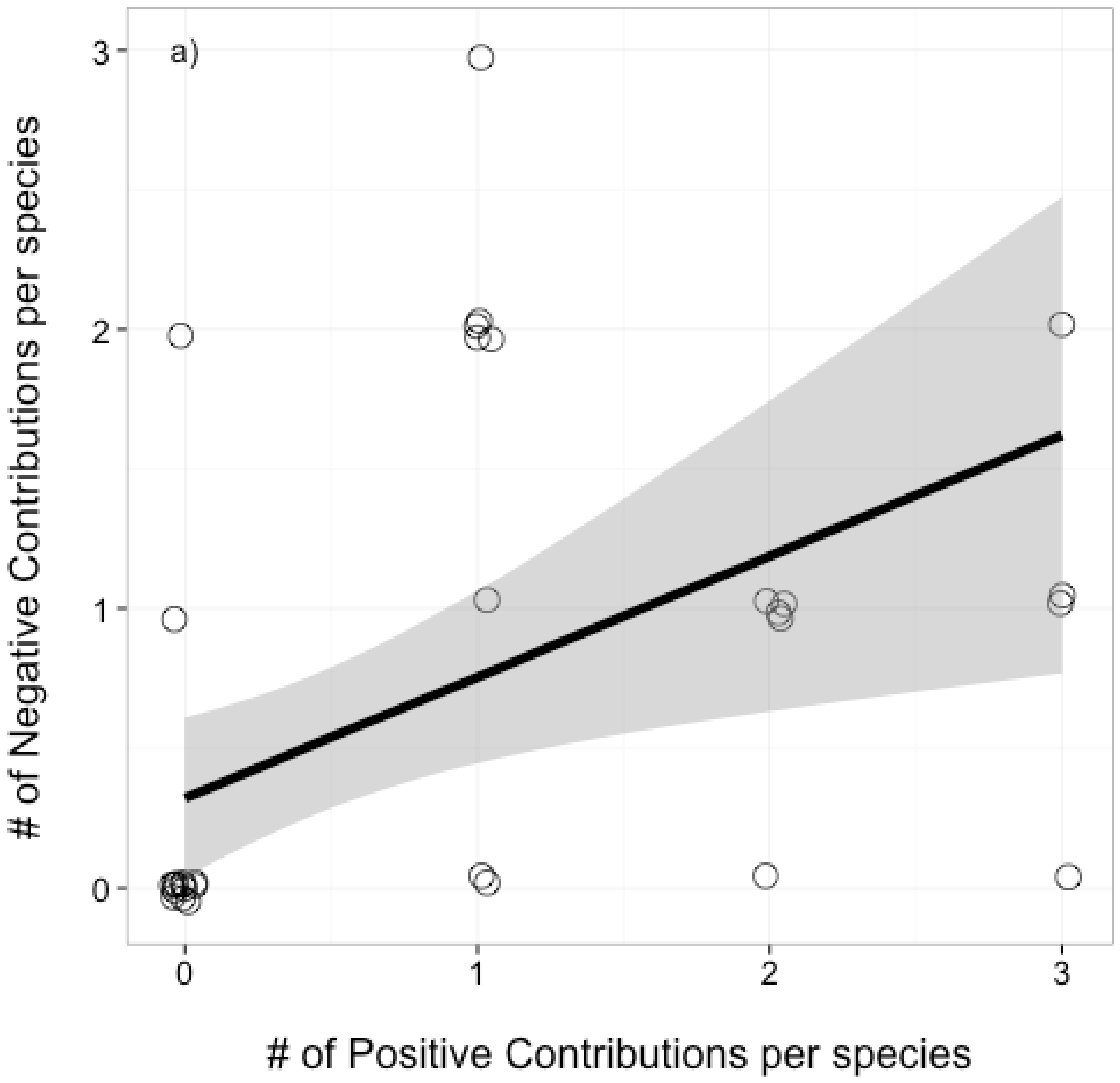

So, a generally positive relationship. We can see from the figure that the slope is slightly less than 1:1, so positive effects accumulate faster. Still, more positive effects = more negative effects. What about the size of those positive versus negative effects?

```
#now get the standardized effect sizes
posNeg<-within(posNeg, {

  stdPosMean <- colSums(filterCoefData(stdCoefs))/Pos
  stdPosMean[which(is.nan(stdPosMean))] <-0

  stdNegMean <- colSums(filterCoefData(stdCoefs, type="negative"))/Neg
  stdNegMean[which(is.nan(stdNegMean))] <-0
})


ggplot(aes(x=stdPosMean,y= stdNegMean), data=posNeg) +
  #  geom_point(size=3) +
  geom_jitter(position = position_jitter(width = .02, height = .02), size=5,
shape=1) +
  theme_bw(base_size=18) +
  xlab("\nAverage Standardized Size of\nPositive Contributions") +
  ylab("Average Standardized Size of\nNegative Contributions\n") +
  stat_smooth(method="lm", colour="black", size=2)+
  annotate("text", x=0, y=0.25, label="b)") +
  geom_abline(slope=-1, intercept=0, size=1, colour="black", lty=3)
```

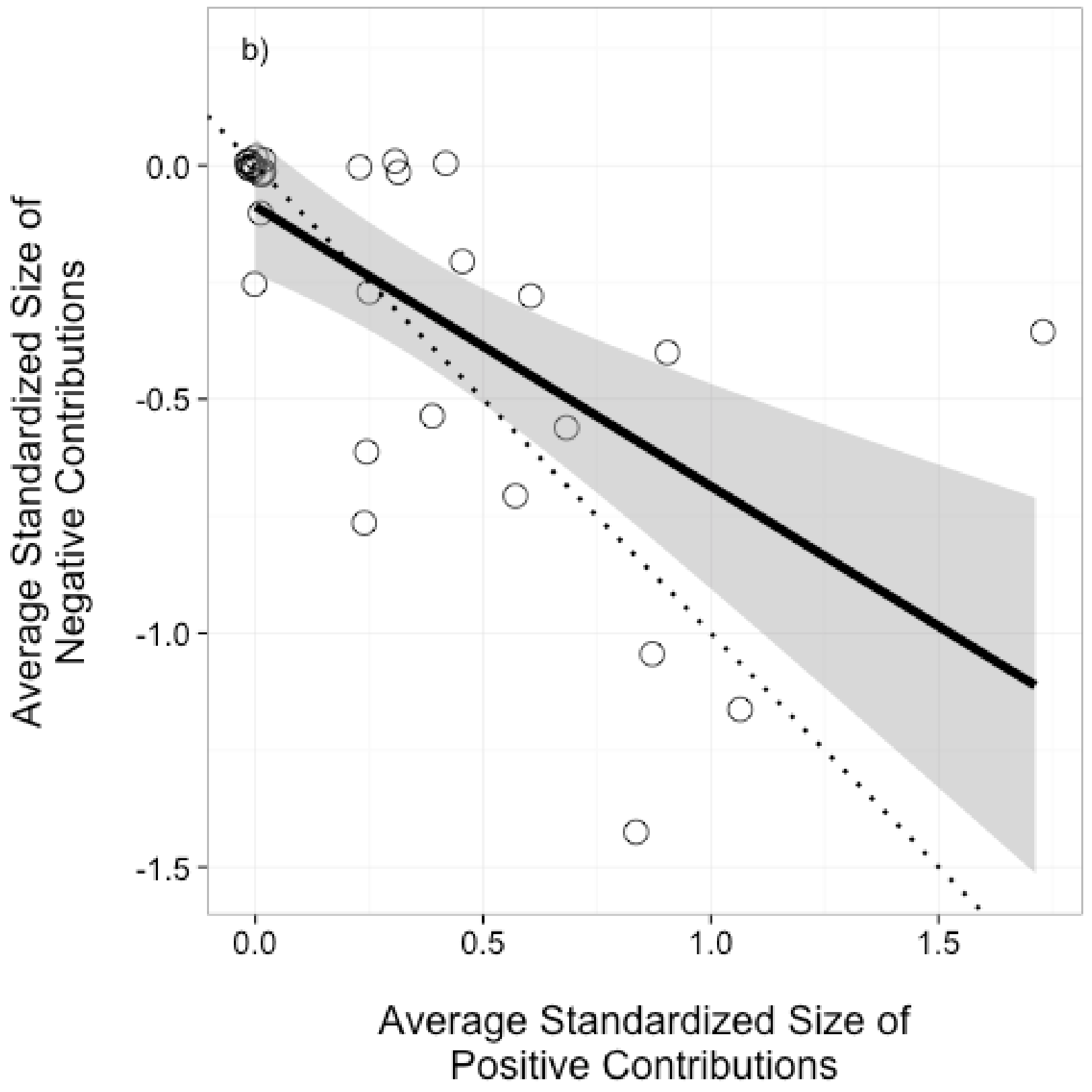

### Averaging Approach

Now we'll try the averaging approach. The first step is to create a set of columns where we have standardized all of the functions of interest, and then create an average of those standardized functions.

```
#add on the new functions along with the averaged multifunctional index
germany<-cbind(germany, getStdAndMeanFunctions(germany, vars))
#germany<-cbind(germany, getStdAndMeanFunctions(germany, vars,
standardizeZScore))
```

We can then plot the averaged multifunctionality quite simply

```
#plot it
ggplot(aes(x=Diversity, y=meanFunction),data=germany)+geom_point(size=3)+
  theme_bw(base_size=15)+
  stat_smooth(method="lm", colour="black", size=2) +
  xlab("\nSpecies Richness") +
  ylab("Average Value of Standardized Functions\n")
```

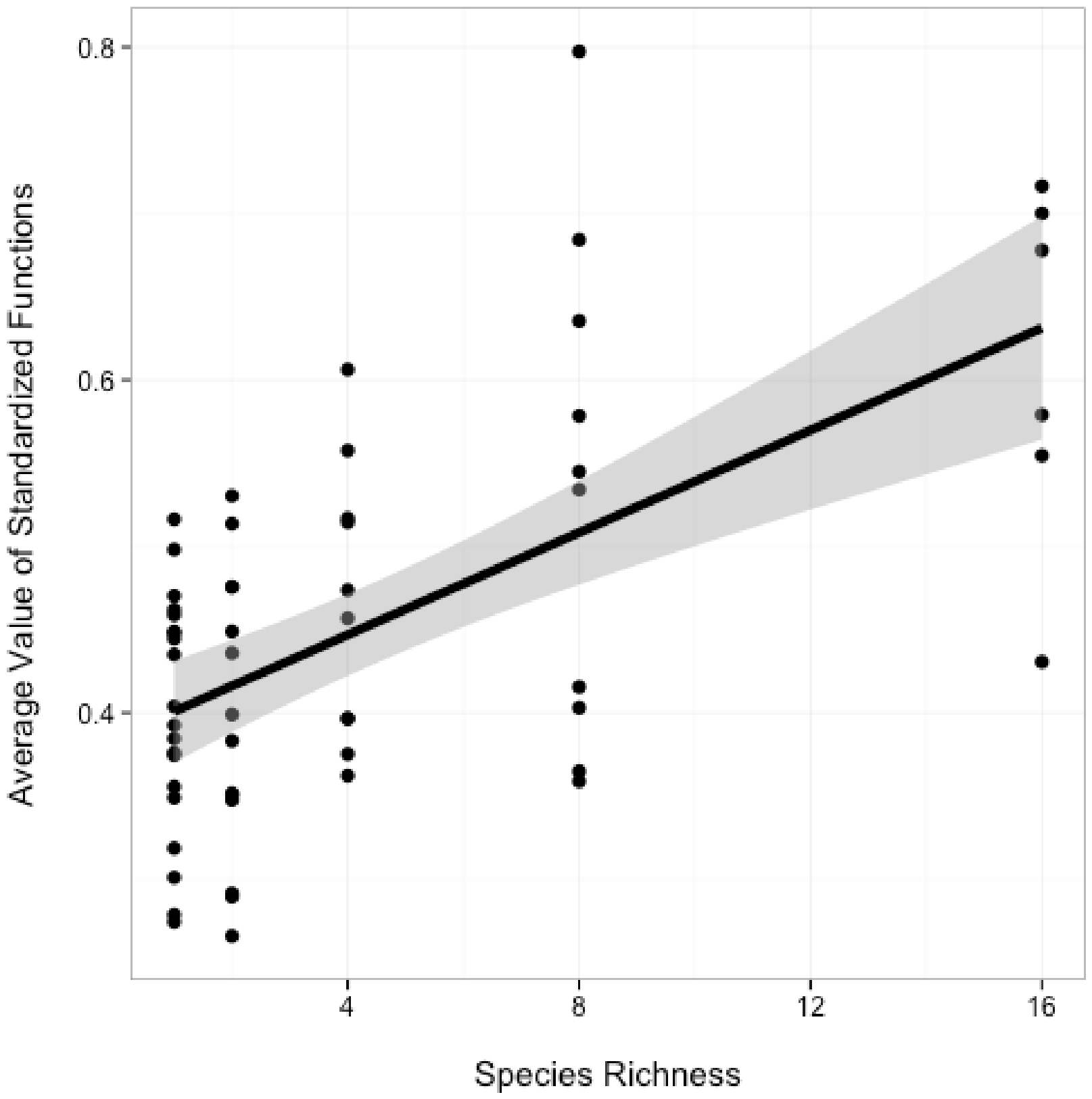

We may want to look at all of the functions on a standardized scale, and add in the averaged line for comparison as well. We can do this by reshaping the data, and plotting it.

```
#reshape for plotting everything with ggplot2
germanyMeanForPlotting<-melt(germany[,c(8,129:134)], id.vars="Diversity")

#nice names for plotting
levels(germanyMeanForPlotting$variable) <- c('Aboveground Biomass', 'Root
Biomass', 'Cotton Decomposition', 'Soil Nitrogen', 'Plant Nitrogen', 'Mean
Multifuncion Index')

#plot it
ggplot(aes(x=Diversity,
y=value),data=germanyMeanForPlotting)+geom_point(size=3)+
  facet_grid(~variable) +
  theme_bw(base_size=15)+
  stat_smooth(method="lm", colour="black", size=2) +
  xlab("\nSpecies Richness") +
  ylab("Standardized Value of Function\n")
```

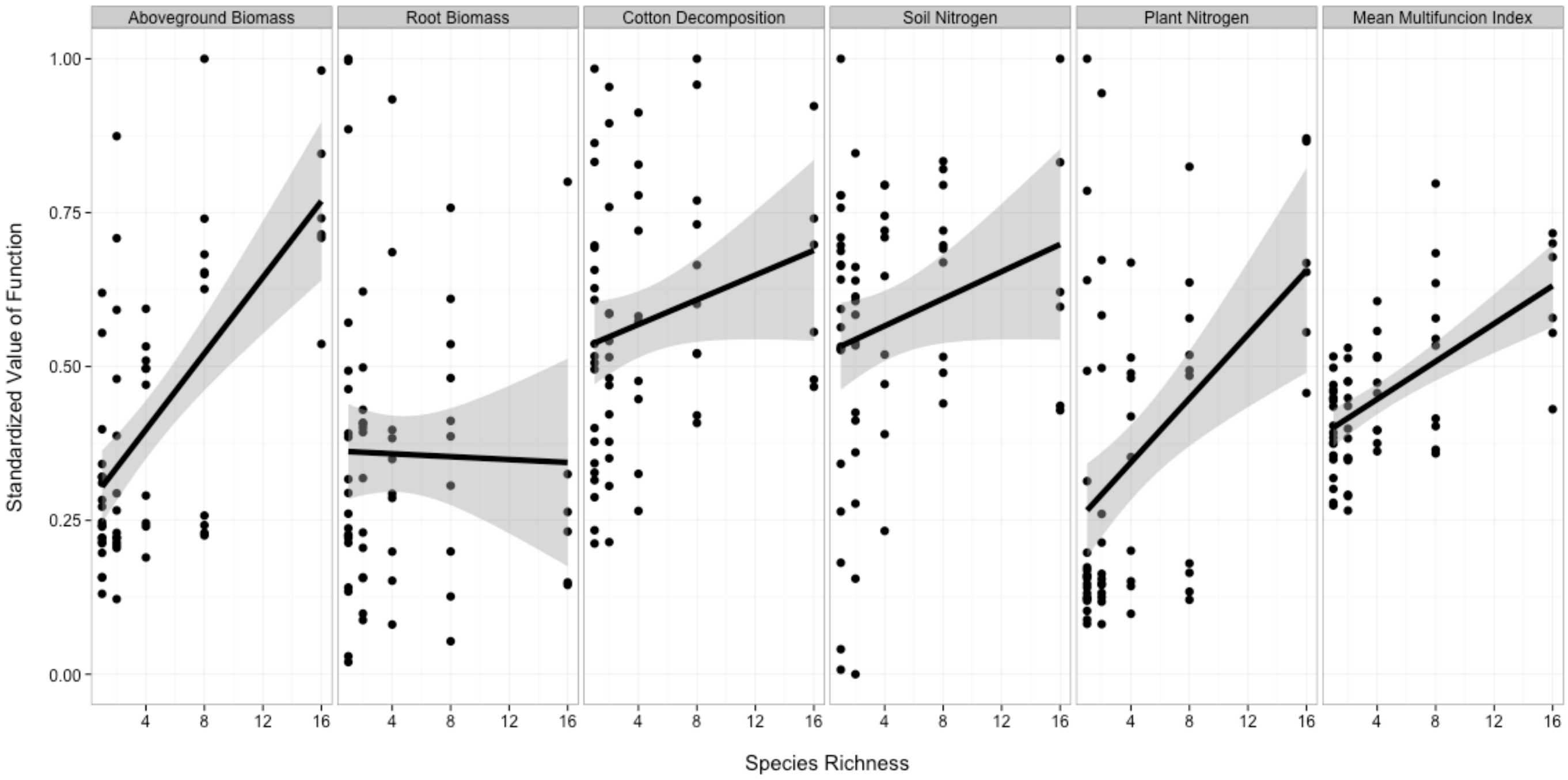

Last, let's test the statistical fit of the effect of diversity on the averaged multifunctionality index.

```
#statistical fit
aveFit<-lm(meanFunction ~ Diversity, data=germany)
Anova(aveFit)
```

```
# Anova Table (Type II tests)
#
# Response: meanFunction
#           Sum Sq Df F value Pr(>F)
# Diversity  0.295  1    32.6  4e-07 ***
# Residuals  0.524 58
# ---
# Signif. codes:  0 '***' 0.001 '**' 0.01 '*' 0.05 '.' 0.1 ' ' 1
```

```
summary(aveFit)
```

```
#
# Call:
# lm(formula = meanFunction ~ Diversity, data = germany)
#
# Residuals:
#      Min       1Q   Median       3Q      Max
# -0.20053 -0.06932  0.00651  0.06272  0.28894
#
# Coefficients:
#             Estimate Std. Error t value Pr(>|t|)
# (Intercept)  0.38542    0.01705   22.61   <2e-16 ***
# Diversity    0.01536    0.00269    5.71    4e-07 ***
# ---
# Signif. codes:  0 '***' 0.001 '**' 0.01 '*' 0.05 '.' 0.1 ' ' 1
#
# Residual standard error: 0.0951 on 58 degrees of freedom
# Multiple R-squared: 0.36, Adjusted R-squared: 0.349
# F-statistic: 32.6 on 1 and 58 DF,  p-value: 4.04e-07
```

**Threshold Approach**

First, we need to create a new data set that looks at the number of functions greater than or equal to a threshold across a wide range of thresholds.

```
germanyThresh<-getFuncsMaxed(germany, vars, threshmin=0.05, threshmax=0.99,
prepend=c("plot","Diversity"), maxN=7)
```

Next, let's perform an analysis on just the 0.8 threshold data.

```
mfuncGermanyLinear08<-glm(funcMaxed ~ Diversity, data=subset(germanyThresh,
germanyThresh$thresholds=="0.8"), family=quasipoisson(link="identity"))

Anova(mfuncGermanyLinear08)
```

```
# Analysis of Deviance Table (Type II tests)
#
# Response: funcMaxed
#           LR Chisq Df Pr(>Chisq)
# Diversity     15.9  1    6.6e-05 ***
# ---
# Signif. codes:  0 '***' 0.001 '**' 0.01 '*' 0.05 '.' 0.1 ' ' 1
```

```
summary(mfuncGermanyLinear08)
```

```
#
# Call:
# glm(formula = funcMaxed ~ Diversity, family = quasipoisson(link =
"identity"),
#     data = subset(germanyThresh, germanyThresh$thresholds ==
#         "0.8"))
#
# Deviance Residuals:
#     Min       1Q   Median       3Q      Max
# -2.1743  -1.1532  -0.0047   0.3821   1.9112
#
# Coefficients:
#             Estimate Std. Error t value Pr(>|t|)
# (Intercept)   0.5517     0.1483    3.72  0.00045 ***
# Diversity     0.1133     0.0331    3.42  0.00116 **
# ---
# Signif. codes:  0 '***' 0.001 '**' 0.01 '*' 0.05 '.' 0.1 ' ' 1
#
# (Dispersion parameter for quasipoisson family taken to be 0.8227)
#
#     Null deviance: 71.852  on 59  degrees of freedom
# Residual deviance: 58.758  on 58  degrees of freedom
# AIC: NA
#
# Number of Fisher Scoring iterations: 4
```

To get a better sense of how robust these results are to threshold choice, let's plot the diversity versus number of functions greater than or equal to a threshold for four different thresholds.

```
gcPlot<-subset(germanyThresh, germanyThresh$thresholds %in% qw(0.2, 0.4, 0.6,
0.8)) #note, using qw as %in% is a string comparison operator

gcPlot$percent<-paste(100*gcPlot$thresholds, "%", sep="")

qplot(Diversity, funcMaxed, data=gcPlot, facets=~percent) +
  stat_smooth(method="glm", family=quasipoisson(link="identity"),
colour="red", lwd=1.2) +
  ylab(expression("Number of Functions" >= Threshold)) +
  xlab("Species Richness") +
  theme_bw(base_size=14) +
  geom_text(data=data.frame(percent = unique(gcPlot$percent),
                       lab = paste(letters[1:4], ")", sep=""),
                       Diversity=2,
                       funcMaxed=6
                       ), mapping=aes(x=Diversity, y=funcMaxed, label=lab))
```

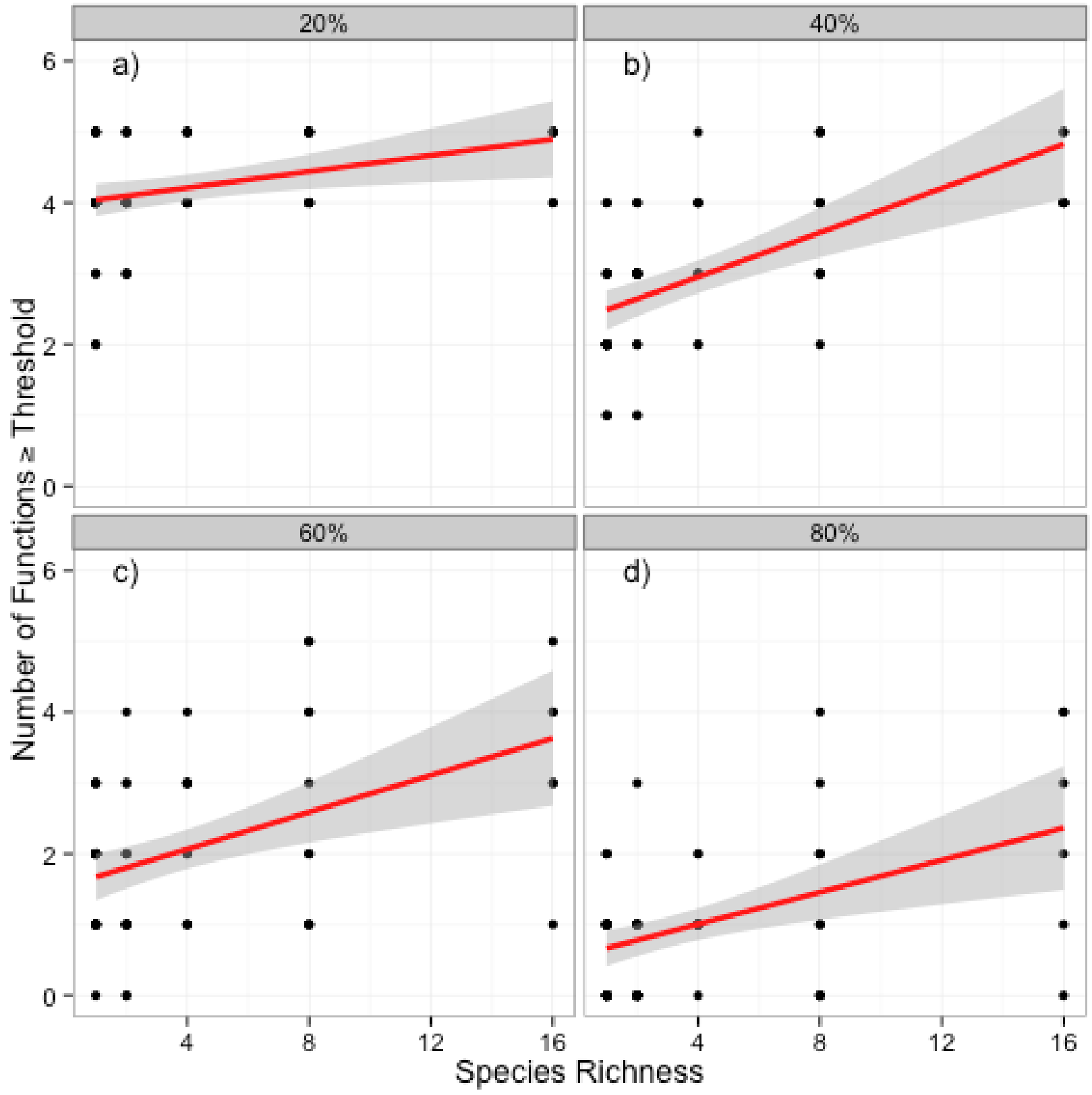

Given that variation, let's look at the entire spread of thresholds.

```
germanyThresh$percent <- 100*germanyThresh$thresholds
ggplot(data=germanyThresh, aes(x=Diversity, y=funcMaxed, group=percent)) +
    ylab(expression("Number of Functions" >= Threshold)) +
    xlab("Species Richness") +
    stat_smooth(method="glm", family=quasipoisson(link="identity"), lwd=0.8,
fill=NA, aes(color=percent)) +
    theme_bw(base_size=14) +
    scale_color_gradient(name="Percent of \nMaximum", low="blue", high="red")
```

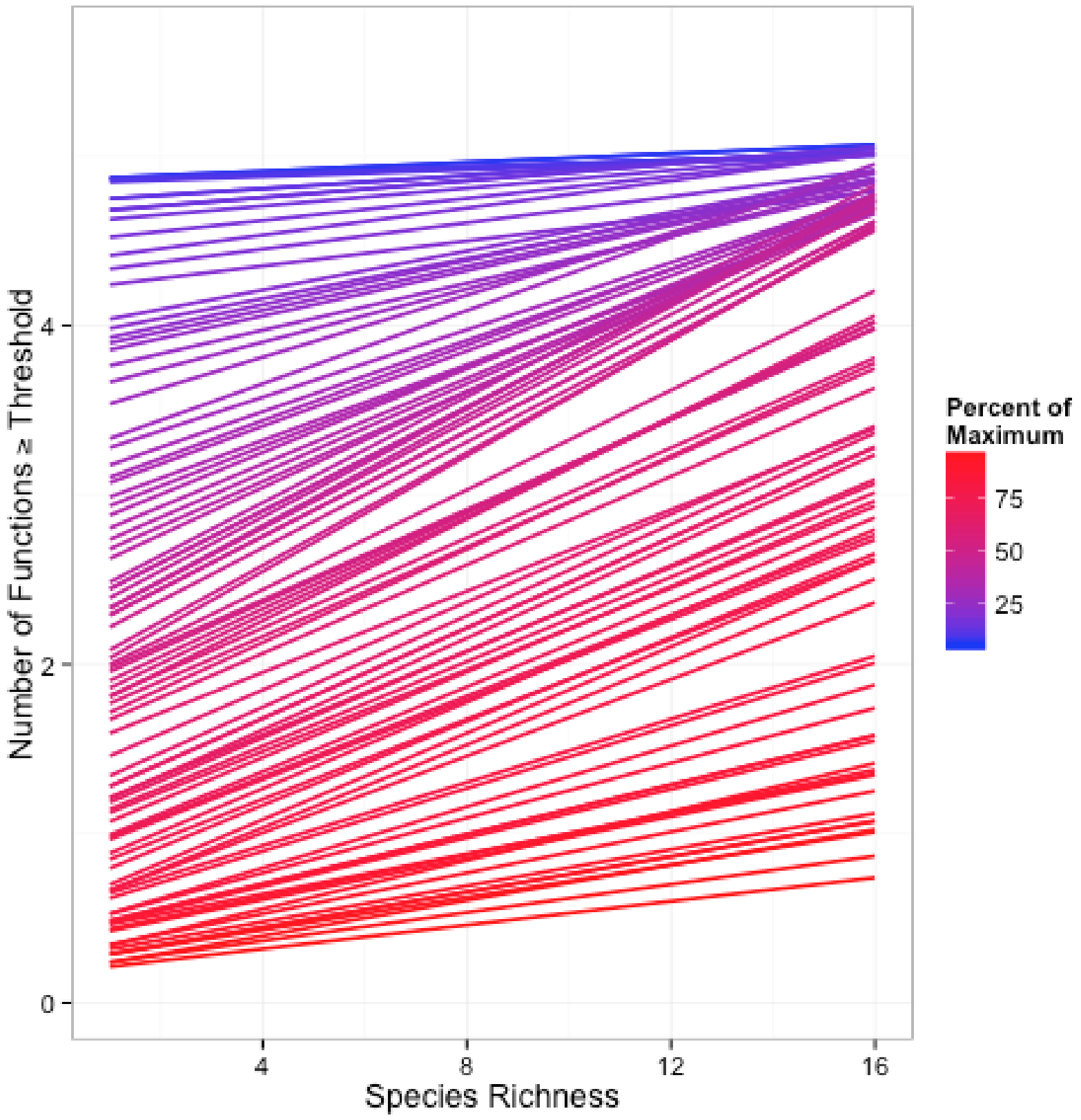

**Multiple Threshold Approach**

To systematically explore the variation in the relationship based on threshold choice, let's look at the slope of the relationship and its confidence interval over all thresholds. We will then plot this relationship.

```
germanyLinearSlopes<-getCoefTab(funcMaxed ~ Diversity, data=germanyThresh,
coefVar="Diversity", family=quasipoisson(link="identity"))


######
# Plot the values of the diversity slope at
# different levels of the threshold
######
germanSlopes <- ggplot(germanyLinearSlopes, aes(x=thresholds)) +
  geom_ribbon(fill="grey50", aes(x=thresholds*100,
ymin=Estimate-1.96*germanyLinearSlopes[["Std. Error"]],

ymax=Estimate+1.96*germanyLinearSlopes[["Std. Error"]])) +
  geom_point(aes(x=thresholds*100, y=Estimate)) +
  ylab("Change in Number of Functions per Addition of 1 Species\n") +
  xlab("\nThreshold (%)") +
  stat_abline(intercept=0, slope=0, lwd=1, linetype=2) +
  theme_bw(base_size=14)

germanSlopes
```

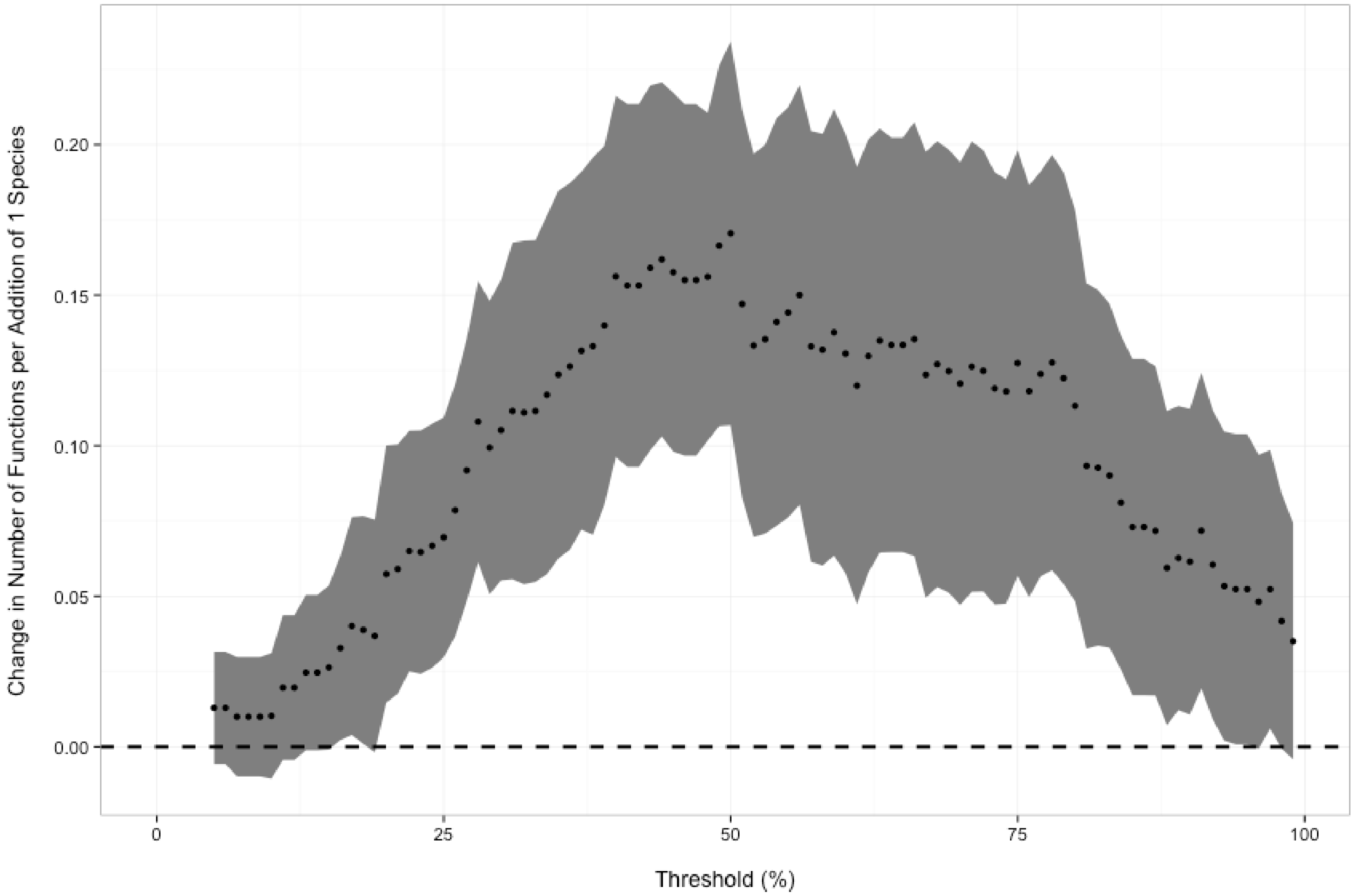

We can easily see a number of indices mentioned in the text - Tmin, Tmax, Tmde, etc. Let's pull them out with our indices function.

```
germanIDX <- getIndices(germanyLinearSlopes, germanyThresh, funcMaxed ~
Diversity)
germanIDX
```

```
#     Tmin Tmax Tmde Rmde.linear Pmde.linear  Mmin   Mmax  Mmde
# 602 0.15 0.98  0.5      0.1705      0.5457 5.021 0.9845 4.474
```

We can now add annotations to our earlier plot to make it more rich, and show us those crucial threshold values.

```
germanyLinearSlopes$Estimate[which(germanyLinearSlopes$thresholds==germanIDX$Tmde)]
```

```
# [1] 0.1705
```

```
germanyThresh$IDX <- 0
germanyThresh$IDX [which(germanyThresh$thresholds %in%
                          c(germanIDX$Tmin, germanIDX$Tmax,
germanIDX$Tmde))] <- 1


ggplot(data=germanyThresh, aes(x=Diversity, y=funcMaxed, group=percent)) +
  ylab(expression("Number of Functions" >= Threshold)) +
  xlab("Species Richness") +
  geom_smooth(method="glm", family=quasipoisson(link="identity"),
              fill=NA, aes(color=percent, lwd=IDX)) +
  theme_bw(base_size=14) +
  scale_color_gradient(name="Percent of \nMaximum", low="blue", high="red") +
  scale_size(range=c(0.3,5), guide="none") +
  annotate(geom="text", x=0, y=c(0.2,2,4.6), label=c("Tmin", "Tmax", "Tmde"))
+
  annotate(geom="text", x=16.7, y=c(germanIDX$Mmin, germanIDX$Mmax,
germanIDX$Mmde), label=c("Mmin", "Mmax", "Mmde"))
```

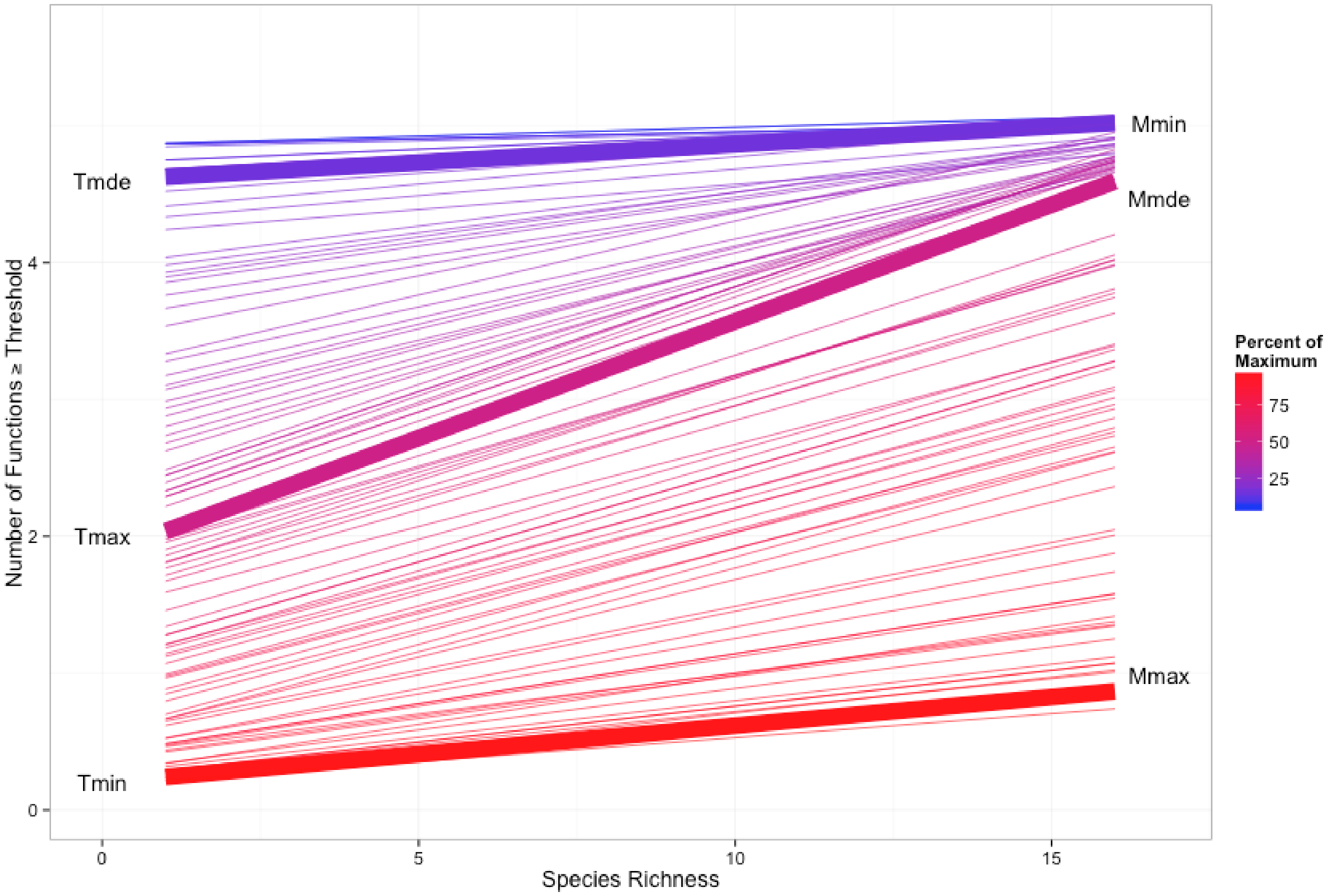

Or we can see where they are on the curve of slopes at different thresholds.

```
germanSlopes + annotate(geom="text", y=c(-0.01, -0.01, -0.01,
germanIDX$Rmde.linear+0.02), x=c(germanIDX$Tmin*100, germanIDX$Tmde*100,
germanIDX$Tmax*100, germanIDX$Tmde*100),  label=c("Tmin", "Tmde", "Tmax",
"Rmde"), color="black")
```

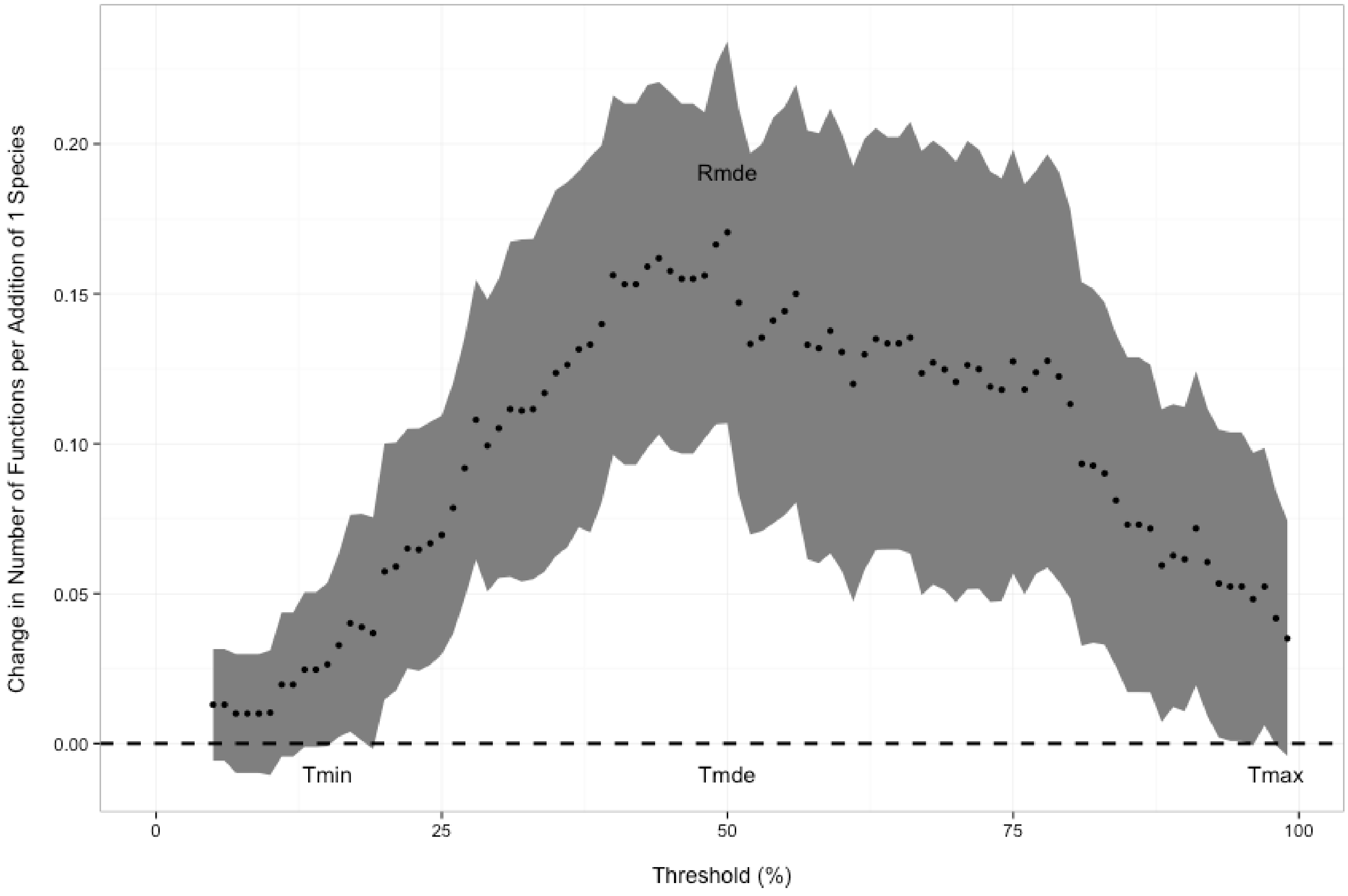

**Comparison of Sites**

Furthermore, we can extend this to the more of the BIODEPTH dataset to compare the strength of multifunctionality across sites. We begin by loading in the data and calculating slopes at different thresholds. We'll subset down to Sheffield, Portugal, and Sweden for simplicity, but this could be done for all sites quite easily.

```
####
# Now we will look at the entire BIODEPTH dataset
#
# Note the use of ddply from plyr to generate the new data frame using all
locations from biodepth.
# If you're not using plyr for your data aggregation, you should be.
http://plyr.had.co.nz/
# It will save you a lot of time and headaches in the future.
####

#Read in data  to run sample analyses on the biodepth data
all_biodepth<-read.table("./FunctContrastData.txt",header=T)
sub_biodepth<-subset(all_biodepth, all_biodepth$location %in% c("Sheffield",
"Portugal", "Sweden"))
sub_biodepth$location<-factor(sub_biodepth$location, levels=c("Portugal",
"Sweden", "Sheffield"))

allVars<-qw(biomassY3, root3, N.g.m2,  light3, N.Soil, wood3, cotton3)
varIdx<-which(names(sub_biodepth) %in% allVars)



#re-normalize so that everything is on the same sign-scale (e.g. the maximum
level of a function is the "best" function)
sub_biodepth<-ddply(sub_biodepth, .(location), function(adf){
  adf$light3<- -1*adf$light3+max(adf$light3, na.rm=T)
  adf$N.Soil<- -1*adf$N.Soil +max(adf$N.Soil, na.rm=T)

  adf
})


#get thresholds
bdThreshes<-ddply(sub_biodepth, .(location), function(x) getFuncsMaxed(x,
vars=allVars, prepend=c("plot","Diversity"), maxN=8))


####look at slopes

#note, maxIT argument is due to some fits needing more iterations to converge
bdLinearSlopes<-getCoefTab(funcMaxed ~ Diversity, data=bdThreshes,
groupVar=c("location", "thresholds"),
                           coefVar="Diversity",
family=quasipoisson(link="identity"), control=list(maxit=800))
```

Great, now we can look at the indices of these three new sites. We'll add Germany in so we can compare them against the German values.

```
indexTable <- lapply(levels(bdLinearSlopes$location), function(x){
  slopedata <- subset(bdLinearSlopes, bdLinearSlopes$location==x)
  threshdata <- subset(bdThreshes, bdThreshes$location==x)
  ret <- getIndices(slopedata, threshdata, funcMaxed ~ Diversity)
  ret<-cbind(location=x, ret)
  ret
})
indexTable <- ldply(indexTable)

indexTable <- rbind(data.frame(location="Germany", germanIDX), indexTable)


indexTable
```

```
#       location Tmin Tmax Tmde Rmde.linear Pmde.linear  Mmin   Mmax  Mmde
# 602   Germany 0.15 0.98 0.50      0.1705      0.5457 5.021 0.9845 4.474
# 2    Portugal 0.09 0.49 0.32      0.2177      0.5080 6.097 4.1058 6.441
# 3      Sweden   NA 0.73 0.46      0.2677      0.5353    NA 2.2013 4.776
# 4   Sheffield 0.58 0.95 0.83      0.1946      0.5839 4.260 2.7135 3.890
```

Finally, let's compare a visualization of the relevant curves at different thresholds to the relationship between threshold and slope of curve to give the table of numbers some visual meaning.

```
library(gridExtra)

bdCurvesAll<-qplot(Diversity, funcMaxed, data=bdThreshes, group=thresholds,
alpha=I(0)) +
  facet_wrap(~location)+#, scales="free") +
  scale_color_gradient(low="blue", high="red", name="Proportion of
\nMaximum", guide=FALSE) +
  stat_smooth(method="glm", lwd=0.8, fill=NA,
family=gaussian(link="identity"), control=list(maxit=200),
aes(color=thresholds)) +
  ylab("\nNumber of Functions ≥ Threshold\n\n") +
  xlab("Species Richness") +
  theme_bw(base_size=15)  +
  geom_text(data=data.frame(location = levels(bdThreshes$location), lab =
paste(letters[1:3], ")", sep=""), thresholds=c(NA, NA, NA)), x=1, y=7,
mapping=aes(label=lab))



#Plot it!
slopePlot<-ggplot(bdLinearSlopes, aes(x=thresholds, y=Estimate)) +
  geom_ribbon(fill="grey50", aes(
x=thresholds*100,ymin=Estimate-2*bdLinearSlopes[["Std. Error"]],
                                  ymax=Estimate+2*bdLinearSlopes[["Std.
Error"]])) +
  geom_point(aes(x=thresholds*100, y=Estimate)) +
  ylab("Change in Number of Functions \nper Addition of 1 Species\n") +
  xlab("Threshold (%)") +
  facet_wrap(~location)+#, scale="free") +
  stat_abline(intercept=0, slope=0, lwd=0.6, linetype=2) +
  theme_bw(base_size=15)+
  geom_text(data=data.frame(location = levels(bdThreshes$location),
                            lab = paste(letters[4:6], ")", sep=""),
                            thresholds=c(NA, NA, NA)), x=0.05, y=0.27,
mapping=aes(label=lab))


###Plot them in a single figure
grid.arrange(bdCurvesAll,
             slopePlot)
```

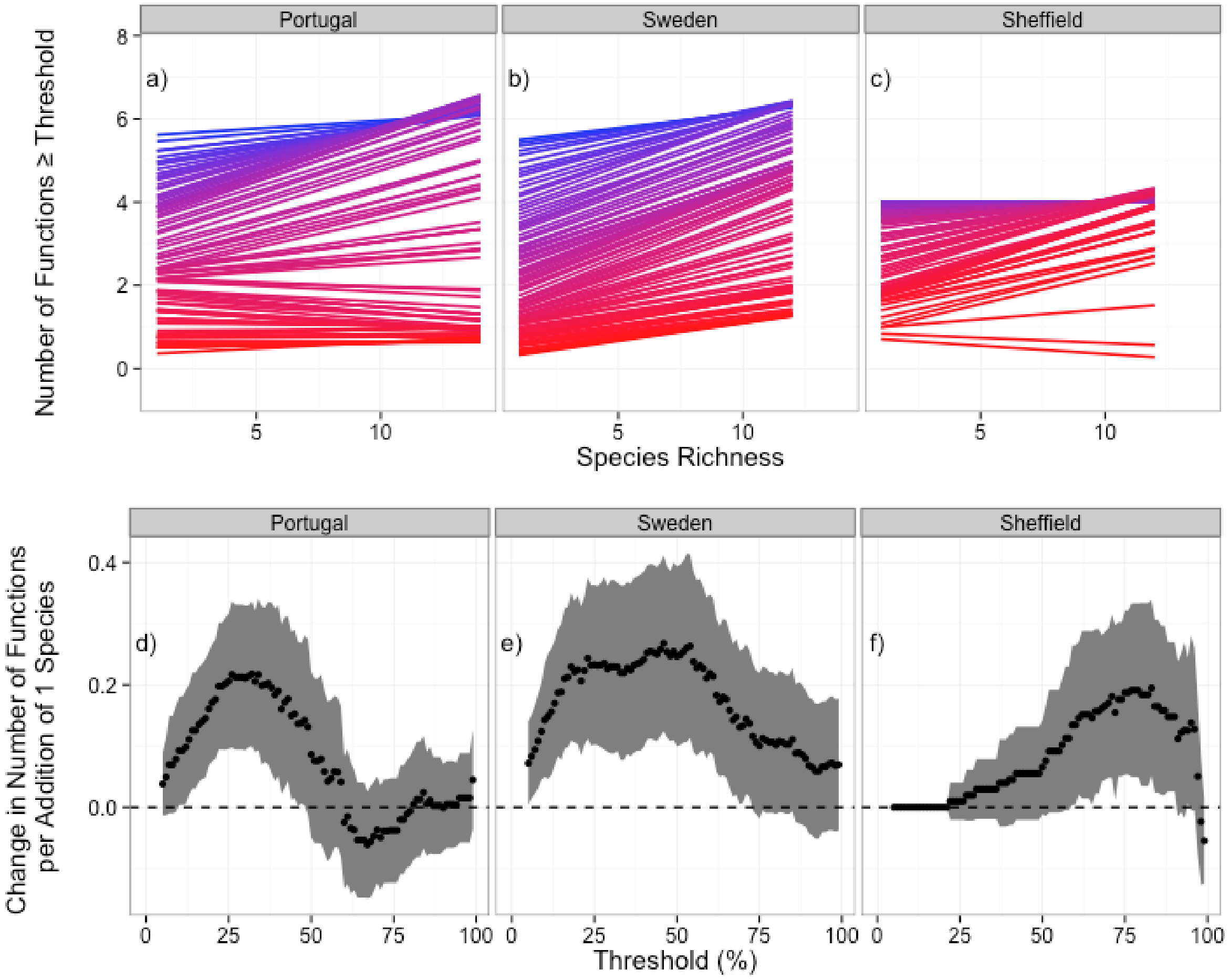

Portugal
Sweden
Sheffield
a)
b)
c)
Number of Functions ≥ Threshold
Species Richness
d)
e)
f)
Change in Number of Functions per Addition of 1 Species
Threshold (%)

**Supplementary Information 2: Choice of underlying model for threshold approach**

We chose a quasipoisson error distribution as 1) we are dealing with count data (O'Hara & Kotze 2010) and 2) the data are likely to be overdispersed (and if not, the dispersion coefficient should converge to 0). We eschewed a negative binomial error, as we had no expectation that the variance around our fitted relationship is skewed (Ver Hoef & Boveng 2007). We chose a linear model for the ease of interpretation of the slope coefficient – the change in the number of functions greater than the threshold per change in species richness. Taking its inverse, we have an estimate of the number of species needed per function. However, we could have fit our model with a log link function, in which case the coefficient for species richness can be interpreted by exponentiating it and subtracting one to assess the change in percent of functions greater than the threshold per change in species richness. The log link also has the advantage of never predicting values of number of functions less than 0. The curves generated by the two fit quite similarly (the difference in deviance of the two models is ~1), so for the purposes of this example, we will continue discussing the linear model. Finally, we note that one could have divided the number of functions greater than or equal to the threshold by the total number of functions measured to estimate the proportion of functions greater than the threshold. In this case, we would then fit a logit curve with a binomial error (Warton 2011). However, this method may be misinterpreted if all relevant functions in an ecosystem have not been measured (most likely in any real research situation). Conversely, if the set of functions measured are chosen for management purposes and the goal of the analysis is to examine the relationship between biodiversity and just that one set of functions are interest, then a logit approach may indeed be appropriate.